\documentclass[aps,twocolumn,letterpaper,superscriptaddress,notitlepage]{revtex4-1}
\usepackage{CJK}
\usepackage{amsmath}
\usepackage{comment}
\usepackage{eufrak}
\usepackage{natbib}
\usepackage{hyperref}
\usepackage[left=0.75in, right=0.7in, top=0.7in, bottom=1in]{geometry}
\usepackage{xcolor}

\begin{document}
\begin{CJK*}{GB}{}
\title{ Magnetoelectric polarizability and optical activity: 
spin and frequency dependence }
\author{Alistair H. Duff$^{1}$ and J.E. Sipe}
\affiliation{Department of Physics, University of Toronto, Ontario M5S 1A7, Canada}

\begin{abstract}
    We extend a microscopic theory of polarization and magnetization to include the spin degree of freedom of the electrons, introducing a general spin orbit coupling and Zeeman interaction term in the Hamiltonian. At finite frequencies and including spin, the magnetoelectric polarizability tensor is replaced by two separate tensors, one that relates the polarization \textbf{P} to the magnetic field \textbf{B} and a separate tensor that relates the magnetization \textbf{M} to the electric field \textbf{E}. When combined with other relevant response tensors a third rank tensor that relates the induced current density to gradients in the electric field can be introduced; it is gauge invariant, in a form suitable for numerical calculations, and describes optical activity -- including spin effects -- even in materials that may lack time reversal symmetry.
\end{abstract}

\maketitle
\end{CJK*}

\section{Introduction}

In considering the linear response of an insulator to static and uniform electric and magnetic fields, one usually focuses on the response of the polarization \textbf{P} to the electric field, and perhaps on that of the magnetization \textbf{M} to the magnetic field.
However, one can also ask under what circumstances there can be a response of \textbf{P} to \textbf{B} and of \textbf{M} to \textbf{E}, where \textbf{B} and \textbf{E} are respectively the Maxwell magnetic and electric fields. These responses are governed by the same ``magnetoelectric polarizability" tensor $\alpha^{il}$ \cite{OMP1,OMP2,OMP3,OMP4,VanderbiltOMP,BerryVanderbilt},
\begin{equation}
\label{MPDefn}
\begin{split}
    P^i = \alpha^{il} B^l,
    \\
    M^i = \alpha^{li} E^l.
\end{split}
\end{equation}
The definitive expression of the magnetoelectric polarizability
tensor for an insulating crystal with spinless electrons in the independent particle approximation, and where the set of occupied bands is topologically trivial, was derived using the ``modern theory of polarization and magnetization" \cite{Macro_Polar,Elec_Polar}. Here polarization and magnetization fields are introduced from a macroscopic perspective, and at most adiabatic variations to the Hamiltonian through the application of uniform fields are considered. The tensor $\alpha^{il}$ consists of two terms, a Chern-Simons contribution that does not lead to any bulk induced charge or current density from the induced polarization and magnetization of \ref{MPDefn}, and a ``cross-gap" term that vanishes unless both time reversal and inversion symmetry are broken. A later extension dealt with the inclusion of contributions to $\alpha^{il}$ from the spin of electrons \cite{SpinMagneto_Adiabatic}.    

More recently, a microscopic approach to defining polarization and magnetization was introduced, where spatially varying and time dependent fields can be considered \cite{Perry_Sipe}. Finite frequency analogs of (\ref{MPDefn}) could be derived,

\begin{equation}
\label{FiniteFreqMP}
\begin{split}
    P^i(\omega) = \alpha^{il}_{\mathcal{P}}(\omega) B^{l}(\omega),
    \\
    M^{i}(\omega) = \alpha^{li}_{\mathcal{M}}(\omega) E^l(\omega),
\end{split}
\end{equation}
but now
\begin{equation}
\label{MPineq}
    \alpha^{il}_\mathcal{P}(\omega) \neq \alpha^{il}_{\mathcal{M}}(\omega).
\end{equation}
This result is not surprising, as the equality of the polarization and magnetization response only holds in the absence of dissipation and dispersion \cite{VanderbiltOMP}. The thermodynamic arguments leading to the equality are only valid in equilibrium, so at arbitrary frequency the equality is broken \cite{OpticalConductivityIvo}.
Further, if terms such as (\ref{FiniteFreqMP}) are included in the optical response, where the right-hand side of the first of (\ref{FiniteFreqMP}) is associated with the dipole moment per unit volume, other
contributions to the induced current density
must also be included. With 
these identified, the linear response of the full macroscopic current density $J^{i}(\textbf{q},\omega)$, where \textbf{q} indicates the wavevector, takes the form

\begin{equation}
\label{CurrentOpto}
J^i(\textbf{q},\omega) = \sigma^{il}(\omega) E^l(\omega) + \sigma^{ilj}(\omega) E^l(\omega) q^j + ...
\end{equation}
where, since the linear response of \textbf{M} to \textbf{B} is usually negligible in the optical regime, $\sigma^{il}(\omega)$ only contains contributions from the frequency-dependent dielectric tensor, while $\sigma^{ilj}(\omega)$ contains contributions from the magnetoelectric polarizability 
tensors $\alpha^{il}_\mathcal{P}(\omega)$ and $\alpha^{il}_{\mathcal{M}}(\omega)$, as well as from the tensor characterizing the response of the quadrupole moment per unit volume to the Maxwell electric field, and from the tensor characterizing the response of the dipole moment per unit volume to the symmetrized derivative of the Maxwell electric field. The full third rank tensor $\sigma^{ilj}(\omega)$ describes the ``optical activity" of the crystal, both its optical rotary dispersion and its circular dichroism \cite{OpticalConductivityIvo}. 

But these calculations were done for spinless electrons \cite{PerryOptical,PerryMagneto}. In this paper we generalize that work to include the contributions to both $\sigma^{il}(\omega)$ and $\sigma^{ilj}(\omega)$ due to the spin dynamics. Since we allow for broken time-reversal symmetry in the unperturbed crystal, $\sigma^{il}(\omega)$ need not be symmetric, reflecting a kind of ``internal Faraday effect," and there are contributions to $\sigma^{ilj}(\omega)$ above and beyond those that are responsible for ``natural optical activity," a term used to describe optical activity in a medium with time-reversal symmetry. As found in the earlier spinless electron calculations, there is a ``gauge invariance" in the expressions for $\sigma^{il}(\omega)$ and $\sigma^{ilj}(\omega)$ that reflects the insensitivity of the results to the choice of the phases of the Bloch functions, and the approach we use avoids the appearance of any ``artificial divergences" that can arise in more standard minimal coupling calculations, and which need to be eliminated by the use of sum rules not always easy to identify \cite{OpticalSipeGhahramani}. Further, we show that our results for $\sigma^{ilj}(\omega)$ (as well as those for $\sigma^{il}(\omega)$) involve only the off-diagonal components of the non-Abelian Berry connection, and hence are particularly suitable for numerical calculations.

In Section \ref{MicroFormalism} we begin by introducing the Hamiltonian and its associated charge and current density operators. We make the frozen ion approximation for the nuclei, and the independent particle approximation for the electrons. The second of these involves the neglect of interactions between the electrons, except as they can be included by employing an effective potential due to the nuclei, and by understanding the ``applied" electromagnetic field we introduce to be the macroscopic Maxwell electromagnetic field \cite{MaxwellFields}.  Then in calculating the dynamical response to such an electromagnetic field we can rely on the equal-time, single particle Green function. We use a generalized Peierls substitution to introduce a ``global Green function," in which no special point has been introduced; if it were appropriate to identify stable units and special points, we would recover the Power-Zienau-Woolley (PZW) transformation \footnote{For a review and references to original work see C. CohenTannoudji, J. Dupont-Roc, and G. Grynberg, Photons
and Atoms. Introduction to Quantum Electrodynamics,
John Wiley and Sons, Inc., 1989} \cite{HealyPZW}. Expanding the Green function in a basis of Wannier functions with orbital type and lattice site labels,
we can then associate 
quantities such as polarization, magnetization, charge density, and current density with the vectors \textbf{R} identifying the lattice sites. These sites act as natural expansion points when developing a multipole expansion.
In this paper we focus on zero temperature insulators for which exponentially localized Wannier functions can be constructed, and outline how the site quantities are constructed, highlighting the modifications that must be made from earlier work \cite{Perry_Sipe} in which the spin-full nature of the particles was neglected.

In Section \ref{MultipoleExpansion}, to prepare for spatial averaging of the microscopic quantities we expand the site polarization fields up to the electric quadrupole moment contribution, and the magnetization fields to the dipole moment contribution. We also give the ground state expressions for the three forms of magnetization identified in this formalism: ``atomic," ``itinerant," and ``spin."  We highlight two explicit changes due to the presence of spin: The first is  due to a modification of the current density by spin-orbit coupling, and the second is to include the intrinsic spin magnetic moment. 

In Section \ref{LinearResponse} a perturbative expansion of the response of the electric and magnetic dipole moments to applied fields is determined, generalizing earlier work where spin was neglected \cite{PerryOptical,PerryMagneto}. We outline how from these we can obtain the macroscopic response tensors, and how the various contributions combine to obtain the optical conductivity tensors. We then show the new spin dependent response of the polarization to an applied magnetic field, and of the spin magnetization to an applied electric field. In the zero-frequency limit these responses are described by the same tensor and are part of the total magnetoelectric polarizability
tensor (MP).

In Section \ref{FreqMP} we discuss how the MP tensor consists of the orbital magnetoelectric polarizability
tensor (OMP) \cite{BerryVanderbilt,OMP2,OMP3,OMP4,VanderbiltOMP}, which can be divided into the usual Chern-Simons and cross-gap contributions, and an explicitly spin dependent contribution, all now generalized to arbitrary frequency. We find the symmetry of the response obtained from the polarization response to a magnetic field and the magnetization to an electric field is lost at finite frequencies. Additionally, the response does not necessarily vanish for time-reversal symmetric systems.  

In Section \ref{OpticalConductivitySection} we combine our results with that of earlier work \cite{PerryOptical} to obtain expressions for the optical conductivity tensors $\sigma^{il}(\omega)$ and $\sigma^{ilj}(\omega)$. We have found that while the expressions that went into creating $\sigma^{ilj}(\omega)$ involve diagonal matrix elements of the Berry connection there is cancellation between the different contributions, and so have written out the tensor in such a way as to make this explicit.

In Section \ref{Conclusions} we summarize the modifications to the formalism to include the spin degree of freedom, the added contribution to the frequency dependent magneto-polarizability. 

We relegate some technical details to Appendices \ref{AppendixA}, \ref{AppendixB}, \ref{AppendixC} and \ref{AppendixD}.

\section{Microscopic Formalism}\label{MicroFormalism}
\subsection{Hamiltonian}

The electronic response of a crystalline insulator is a consequence of the evolution of the fermionic electron field operator, $\hat{\psi}(\textbf{x},t)$. In the Heisenberg picture the dynamics are governed by

\begin{equation}
    i\hbar \frac{\partial \hat{\psi}(\textbf{x},t)}{\partial t} = \mathcal{H}(\textbf{x},t) \hat{\psi}(\textbf{x},t),
\end{equation}
where $\hat{\psi}(\textbf{x},t)$ is a two-component Pauli spinor operator. We make the frozen ion and independent particle approximations, and begin by taking the electrons to be subject to 
an effective potential energy term $\text{V}(\textbf{x})$ that has the same periodicity as the crystal lattice, $\text{V}(\textbf{x}) = \text{V}(\textbf{x}+\textbf{R})$ for all Bravais lattice vectors \textbf{R}. 
We can then include
relativistic corrections that involve the spin by taking the differential operator for the Hamiltonian to be  
\begin{equation}
    \mathcal{H}^0_{TRS}(\textbf{x}) = -\frac{ \hbar^2}{2m} \nabla^2 +\text{V}(\textbf{x})+\mathcal{H}^0_{SOC}(\textbf{x}).
\end{equation}
$\mathcal{H}^0_{SOC}$ is the spin-orbit coupling term  \cite{FoldyDirac,Bjorken_Drell,SchiffDirac,BerryPhase}, which we write in a general form as

\begin{equation}
    \mathcal{H}^0_{SOC}(\textbf{x}) = -\frac{i\hbar^2}{4m^2c^2}\boldsymbol\sigma\cdot\nabla \text{V}(\textbf{x})\times\nabla,
\end{equation}
where $\boldsymbol\sigma$ is the vector of Pauli matrices that act on spinor wavefunctions. Other relativistic corrections, such as the Darwin and mass-velocity terms \cite{Bjorken_Drell}, are neglected. 

We use the subscript ${TRS}$ to indicate that $\mathcal{H}^0_{TRS}(\textbf{x})$ satisfies time-reversal symmetry. 
Even before any electromagnetic fields are applied, we can allow for this symmetry to be broken by introducing a possible
 ``internal", static, cell-periodic vector potential \textbf{A}$_\text{static}$(\textbf{x}), where \textbf{A}$_\text{static}(\textbf{x})$ = \textbf{A}$_\text{static}(\textbf{x}+\textbf{R})$. While the presence of $\textbf{A}_\text{static}(\textbf{x})$ breaks time-reversal symmetry, its inclusion does not break the translational symmetry of the Hamiltonian, and thus Bloch's theorem can still be applied. The introduction of such an $\textbf{A}_\text{static}(\textbf{x})$ is a common approach to including the effects of broken time reversal symmetry within a basic independent particle approximation \cite{Haldane_Julen}.   

 We also consider the system to be perturbed by an applied electromagnetic field characterized by a scalar potential $\phi(\textbf{x},t)$ and a vector potential $\textbf{A}(\textbf{x},t)$. We treat the applied electromagnetic field and the internal magnetic field classically, and adjust the Hamiltonian for their inclusion following the standard minimal coupling prescription, as well as adding in the magnetic dipole energy term due to the spin. Defining
\begin{equation}
    \mathfrak{p}(\textbf{x}) = -i\hbar \nabla - \frac{e}{c} \textbf{A}_\text{static}(\textbf{x}),
\end{equation}
where we take the charge of the electron to be $e = -|e|$, the full differential operator $\mathcal{H}(\textbf{x},t)$ is then given by

\begin{widetext}
\begin{equation}
\label{scriptH}
\begin{split}
    \mathcal{H}(\textbf{x},t) = \frac{(\boldsymbol{\mathfrak{p}}(\textbf{x})-\frac{e}{c}\textbf{A}(\textbf{x},t))^2}{2m} + \text{V}(\textbf{x})
    +e\phi(\textbf{x},t)-\frac{e\hbar}{2mc} \boldsymbol\sigma\cdot{\textbf{B}}(\textbf{x},t) -\frac{e\hbar}{2mc} \boldsymbol\sigma\cdot\textbf{B}_\text{static}(\textbf{x})
    \\
    +\frac{\hbar}{4m^2c^2}\boldsymbol\sigma\cdot\nabla\text{V}(\textbf{x})\times(\boldsymbol{\mathfrak{p}}(\textbf{x})-\frac{e}{c}\textbf{A}(\textbf{x},t)),
\end{split}
\end{equation}
\end{widetext}
where the applied magnetic field is $\textbf{B}(\textbf{x},t)$ = $\nabla \times \textbf{A}(\textbf{x},t)$ and $\textbf{B}_\text{static}(\textbf{x}) = \nabla\times \textbf{A}_\text{static}(\textbf{x})$. This independent particle treatment, where the breaking of time-reversal symmetry is described only by the introduction of $\textbf{A}_\text{static}(\textbf{x})$, could be extended by including another Zeemen term involving a $\textbf{B}_\text{eff}(\textbf{x})$, nonuniform but  with the periodicity of the lattice, that would capture a mean-field description of exchange effects; see, e.g Kohn and Sham \cite{KohnSham-Ex}, and Ogata \cite{OgataMagnetizationBloch}. We do not explicitly implement that approach here, because we want to simplify the comparison of our results with earlier calculations that were done with the more basic independent particle approximation where such exchange effects are neglected. We plan to turn to such an extension in a later communication.
We assume that the gradient of V(\textbf{x}) is much larger than the applied electric field, in regions of space where that gradient is important,
and therefore the spin-orbit interaction is not modified by the applied electric field at our level of approximation. At times when the applied electromagnetic field vanishes, the expression for $\mathcal{H}(\textbf{x},t)$ reduces to   $\mathcal{H}^0(\textbf{x})$, where the latter is obtained from the former by setting $\phi(\textbf{x},t)$ and $\textbf{A}(\textbf{x},t)$ equal to zero. $\mathcal{H}^{0}(\textbf{x})$ is thus the ``unperturbed Hamiltonian", with $\textbf{A}_\text{static}(\textbf{x})$ and $\textbf{B}_\text{static}(\textbf{x})$ included to allow for more interesting ground state Bloch functions that may break time-reversal symmetry.

The full Hamiltonian is then given by

\begin{equation}
    \hat{H}(t) = \int \hat{\psi}^\dag(\textbf{x},t) \mathcal{H}(\textbf{x},t) \hat{\psi}(\textbf{x},t) d\textbf{x},
\end{equation}
where $\mathcal{H}(\textbf{x},t)$ is a 2x2 matrix with in general four non-zero components. In the usual spin-z basis we can denote the matrix elements by two spin-arrow labels; however, in general we introduce sans-serif subscripts to indicate spinor components, $\mathcal{H}_\textsf{ij}(\textbf{x},t)$.

\begin{widetext}
\subsection{Charge and Current Densities}

The equations for the charge and current densities are obtained \cite{QuantumFieldTheoretic} via 

\begin{equation}
\label{rho}
    \hat{\rho}(\textbf{x},t) = \frac{\delta \hat{H}(t)}{\delta \phi(\textbf{x},t)}+ \rho^\text{ion}(\textbf{x}),
\end{equation}
and
\begin{equation}
\label{j}
    \hat{\textbf{j}}(\textbf{x},t) = -c\frac{\delta \hat{H}(t)}{\delta \textbf{A}(\textbf{x},t)},
\end{equation}
where to the electronic charge density in (\ref{rho}) we have added a charge density for the (assumed fixed) ions so that  $\hat{\rho}(\textbf{x},t)$ is the full charge density. Here the ion charge density $\rho^\text{ion}(\textbf{x})$ can be written as 
\begin{equation}
    \rho^\text{ion}(\textbf{x}) = \sum_\textbf{R} \rho_\textbf{R}^\text{ion}(\textbf{x}),
\end{equation}
where $\rho_\textbf{R}^\text{ion}(\textbf{x})$ is the ion density associated with lattice site $\textbf{R}$.  If the ions are approximated as point charges, we have
\begin{equation}
\label{nucleidensity}
    \rho_\textbf{R}^\text{ion}(\textbf{x}) = \sum_N q_N \delta(\textbf{x}-\textbf{R}- \textbf{d}_N),
\end{equation}
where we assume that in each unit cell there are N ions with charges q$_N$ located at $\textbf{R}+\textbf{d}_N$.

The charge and current densities (\ref{rho}, \ref{j}) satisfy the continuity equation, and are given by
\begin{equation}
    \hat{\rho}(\textbf{x},t) = e\hat{\psi}^\dag(\textbf{x},t)\hat{\psi}(\textbf{x},t)+ \rho^\text{ion}(\textbf{x}),
\end{equation}
and

\begin{equation}
\label{cden}
\begin{split}
    \hat{\textbf{j}}(\textbf{x},t) = \frac{e}{2m}\big( \hat{\psi}^\dag(\textbf{x},t)(\boldsymbol{\mathfrak{p}}(\textbf{x}) - \frac{e}{c}\textbf{A}(\textbf{x},t))\hat{\psi}(\textbf{x},t) 
    + ((\boldsymbol{\mathfrak{p}}(\textbf{x})-\frac{e}{c}\textbf{A}(\textbf{x},t))\hat{\psi}(\textbf{x},t))^\dag\hat{\psi}(\textbf{x},t)\big)
    \\
    + \hat{\textbf{j}}_\textbf{m}(\textbf{x},t)+\frac{e\hbar}{4m^2c^2}\hat{\psi}^\dag(\textbf{x},t)(\boldsymbol\sigma\times\nabla \text{V}(\textbf{x}))\hat{\psi}(\textbf{x},t).
\end{split}
\end{equation}
\end{widetext}

The first line on the right-hand-side of (\ref{cden}) is the standard current density that one would expect from a minimal coupling Hamiltonian. The second line has two additional contributions that are associated with spin: The
first, $\hat{\textbf{j}}_\textbf{m}(\textbf{x},t)$, is the magnetization current  that would be present even for a free electron gas \cite{Griffiths}; it is given by 
\begin{equation}
    \hat{\textbf{j}}_\textbf{m}(\textbf{x},t) = c\nabla\times \hat{\textbf{m}}_{\boldsymbol{\sigma}}(\textbf{x},t),
\end{equation}

\noindent where 

\begin{equation}
    \hat{\textbf{m}}_{\boldsymbol\sigma}(\textbf{x},t) = \frac{e\hbar}{2m c} \hat{\psi}^\dag(\textbf{x},t)\boldsymbol\sigma\hat{\psi}(\textbf{x},t).
\end{equation}
The remaining contribution to the charge current density in (\ref{cden}) is transverse to the spin magnetization of the system and to the electric field $\textbf{E}_\text{lattice}=-e^{-1}\nabla\text{V}$ created by the crystal environment. It arises because the spin-orbit coupling introduces a correction to the relation between the velocity and momentum of an electron \cite{ElectronCurrent,SpinCurrentNano,SpinDynamics,BerryCurvOrb}. Physically, this can be understood as a consequence of the effective electric dipole moment $\boldsymbol\mu =c^{-1} \textbf{v} \times \boldsymbol\nu$ \cite{MovingMagneticDipole} in the laboratory frame of a magnetic dipole moment $\boldsymbol\nu$ moving with velocity
$\textbf{v}$ 
\footnote{The Lagrangian of a single electron interacting with the electromagnetic field, which is specified in the laboratory frame, is then 
\begin{equation}
\label{Lagrangian}
    \text{L}=\frac{1}{2}m\textbf{v}\cdot\textbf{v}+\frac{e}{c}\textbf{v}\cdot(\textbf{A}+\textbf{A}_\text{static})+\boldsymbol\mu\cdot\textbf{E}_\text{lattice},
\end{equation}
together with contributions from the scalar potential and from the interaction of the magnetic dipole moment with the magnetic field. Since the last two terms on the right-hand-side of (\ref{Lagrangian}) both involve the velocity $\textbf{v}$, they will both lead to a difference between the canonical momentum and the product of the mass and the velocity. Putting the spin magnetic moment $\boldsymbol{\nu}=e\textbf{S}/(mc)$, where $\textbf{S}$ is the spin angular momentum, and including in this term the Thomas factor $\frac{1}{2}$ to account for Thomas precession  \cite{ThomasPrecession,JacksonEM}, we find 
\begin{equation}
    \textbf{p}=m\textbf{v}+\frac{e}{c}(\textbf{A}+\textbf{A}_\text{static})-\frac{1}{2mc^2}(\textbf{S}\times\nabla\text{V}),
\end{equation}
Taking $\textbf{S}\rightarrow{\hbar\boldsymbol\sigma}/2$, we can write this equation as 
\begin{equation}
\label{spcden}
    e\textbf{v}=\frac{e}{m}(\textbf{p}-\frac{e}{c}(\textbf{A}+\textbf{A}_\text{static}))+\frac{e\hbar}{4m^2c^2}(\boldsymbol\sigma\times\nabla\text{V}).
\end{equation}
The second term in the second line of the expression (\ref{cden}) for the current density is the field theory analogue of the last term on the right-hand-side of (\ref{spcden}).  
}.

\subsection{Wannier functions}

Our approach employs a Wannier function basis for calculations \cite{CohenCondensedMatter,BerryVanderbilt}. There are various basis transformations that are made in this section, so we begin by giving a motivation for the steps we are about to perform.

The first ``natural" basis for a crystal would be the Bloch eigenstate solutions $|n\textbf{k}\rangle$ of the unperturbed Hamiltonian introduced in section \ref{MicroFormalism} that, within a phase factor, satisfy discrete translational invariance. However, many matrix elements, such as position and angular momentum, must be treated with care in the Bloch basis \cite{SeitzSolid,BerryVanderbilt}. In using a Wannier function basis these matrix elements are well-defined and not plagued by complications involving derivatives of Dirac delta functions, and the like.

There is much freedom in the choice of Wannier functions. This arises not only from the phase indeterminacy of the Bloch functions that then enter in the construction of the Wannier functions, but also because one can apply a general unitary transformation and ``mix" the occupied bands involved in constructing the Wannier functions associated with the ground state. This unitary transformation is leveraged in the construction of ``maximally localized Wannier functions" that have well-behaved localization properties \cite{MaxLocWannier,ExponentialLocWannier}. Thus we will introduce a modified Bloch basis $|\alpha \textbf{k}\rangle$, where the states are not eigenstates of the Hamiltonian. This basis will be an important intermediate step in considering the transformation between matrix elements in the Wannier and original Bloch basis.

To treat a general applied magnetic field we also introduce an ``adjusted" Wannier basis. While the construction and implementation is an added complication it allows for the systematic inclusion of the wavefunction modifications induced by the magnetic field. This is crucial in magnetic response formulae such as that determining the magnetic susceptibility \cite{OgataCorrectionsPeierls}.

Working toward the construction of the ``adjusted" Wannier function basis we begin with the two component spinor wave functions 

\begin{equation}
\label{Blochfunctions}
    \psi_{n\textbf{k}}(\textbf{x}) \equiv \langle \textbf{x} | \psi_{n\textbf{k}} \rangle = \frac{1}{(2\pi)^\frac{3}{2}} e^{i\textbf{k}\cdot \textbf{x}} \begin{bmatrix}
    u_{n\textbf{k},\uparrow}(\textbf{x})
    \\
    u_{n\textbf{k},\downarrow}(\textbf{x})
    \end{bmatrix}, 
\end{equation}
which are the Bloch eigenfunctions of the unperturbed Hamiltonian $\mathcal{H}^0(\textbf{x})$ introduced after equation (\ref{scriptH}). The Bloch eigenfunctions are normalized over the infinite crystal such that $\langle\psi_{m\textbf{k}'}|\psi_{n\textbf{k}}\rangle = \delta_{nm}\delta(\textbf{k}-\textbf{k}')$; when sans-serif spinor component indices are omitted the object is understood to be the full two-component spinor. Associated with each Bloch eigenvector is an energy $E_{n\textbf{k}}$ and a cell-periodic spinor function $u_{n\textbf{k}}(\textbf{x})$. The n are band indices and $\hbar\textbf{k}$ the crystal momentum. The cell-periodic functions $u_{n\textbf{k}}(\textbf{x}) \equiv \langle \textbf{x}|n\textbf{k}\rangle$ satisfy the orthogonality condition $(m\textbf{k}|n\textbf{k}) = \delta_{nm}$, where we adopt the notation that 

\begin{equation}
\label{unitcellintegral}
    (g|h) \equiv \frac{1}{\Omega_{uc}}\int_{\Omega_{uc}} g_{\textsf{i}}^*(\textbf{x})h_{\textsf{i}}(\textbf{x}) d\textbf{x},
\end{equation}
where $\Omega_{uc}$ is the unit cell volume. Repeated spinor indices are implicitly summed over. 

It is useful to introduce sets of Wannier functions, where the Wannier functions in a particular set are associated with a particular set of ``isolated bands" that may intersect amongst themselves, but where no band from the set intersects with bands from different sets. For the Wannier functions associated with a particular set of isolated bands, the prescription for doing that is  

\begin{equation}
\label{WannierKet}
    |\alpha\textbf{R}\rangle = \sqrt{\frac{\Omega_{uc}}{(2\pi)^3}} \int_{BZ} d\textbf{k} e^{-i\textbf{k}\cdot\textbf{R}} \sum_n U_{n\alpha}(\textbf{k}) |\psi_{n\textbf{k}}\rangle,
\end{equation}
 where the unitary matrix U(\textbf{k}) and the Bloch eigenvectors $|\psi_{n\textbf{k}}\rangle$ are chosen to be periodic over the first Brillouin zone. The sum over band indices in (\ref{WannierKet}) is only over the relevant set of isolated bands, and each Wannier function $|\alpha\textbf{R}\rangle$ is labeled by a type index $\alpha$ and a lattice site $\textbf{R}$ with which it is identified.
 
 As a first example, in this paper we consider an insulator, and so we can introduce two sets of isolated bands: One consists of the valence bands, and the other of the conduction bands.  In studies of insulators involving scalar wave functions, if the band structure is ``topologically trivial" then one can construct one set of exponentially localized Wannier functions (ELWF)  \cite{ExponentialLocWannier,MaxLocWannier,SmoothGaugeVanderbilt,SmoothGaugeWannier,BlochBundlesMaxLoc} from the valence bands, and another set from the conduction bands. For the scalar case the restriction is to insulators where the set of bands chosen has net zero Chern invariant; and work has been done to develop a procedure for constructing Wannier functions for Z2 insulators \cite{SmoothGaugeVanderbilt,Z2Insulator}. In the generalization to spinor wave functions developed here we consider the analogous scenario, assuming that we can associate a set of spinor ELWF with the valence bands and another set with the conduction bands.
 
 Even with the restriction to ELWFs, the Wannier functions are not uniquely defined, due to the freedom in choosing the unitary transformations U(\textbf{k}).  Thus there is an unavoidable ``gauge freedom" -- to be distinguished from the gauge freedom in choosing the scalar and vector potentials that identify the electromagnetic field -- when identifying the Wannier functions. In this paper we refer to quantities as being gauge-dependent, in this sense, if they depend on the matrices U(\textbf{k}) or their derivatives. This gauge freedom does not extend to physical quantities such as the charge and current densities. 

The expression (\ref{WannierKet}) for the Wannier functions provides a mapping from the set of filled energy eigenstates to the set of filled ELWFs; likewise, we can map the unoccupied states to unoccupied ELWFs. At zero temperature the ground state filling factors $f_{n}$ for the bands are either 0 or 1. Likewise the orbital filling factors $f_{\alpha}$ are either 0 or 1, and if $U_{n\alpha} \neq 0$ then $f_{n} = f_{\alpha}$.

The Wannier functions $W_{\alpha\textbf{R}}(\textbf{x}) \equiv \langle \textbf{x}| \alpha\textbf{R}\rangle$ form an orthogonal set,

\begin{equation}
    \int W^*_{\beta\textbf{R}',\textsf{i}}(\textbf{x}) W_{\alpha\textbf{R},\textsf{i}}(\textbf{x}) d\textbf{x} = \delta_{\beta\alpha}\delta_{\textbf{R}'\textbf{R}}.
\end{equation}
Again following the strategy for spinless particles, we
introduce a new set of kets
\{$|\alpha\textbf{k}\rangle$\}, with coordinate representations that are cell-periodic, and that are linked to
Wannier functions of type $\alpha$,

\begin{equation}
    |\alpha\textbf{k}\rangle = U_{n\alpha}(\textbf{k})|n\textbf{k}\rangle.
\end{equation}
In the case of a multi-band unitary transformation relating the original set of Bloch functions $\{ |n\textbf{k}\rangle\}$ to the new set $\{ |\alpha\textbf{k}\rangle\}$, the set $\{ |\alpha\textbf{k}\rangle\}$ are not identified with eigenstates of the Hamiltonian. Matrix elements of the position operator between Wannier functions can be related to the non-Abelian Berry connection \cite{SeitzSolid} associated with this new set of cell-periodic functions

\begin{equation}
\begin{split}
\label{BerryConnection}
    \int W^*_{\beta\textbf{R},\textsf{i}}(\textbf{x}) x^a W_{\alpha\textbf{0},\textsf{i}}(\textbf{x}) d\textbf{x} =
    \frac{\Omega_{uc}}{(2\pi)^3} \int_{BZ} d\textbf{k} e^{i\textbf{k}\cdot\textbf{R}} \tilde{\xi}^a_{\beta\alpha}(\textbf{k}),
\end{split}
\end{equation}
where 

\begin{equation}
\label{Non-Abelian}
    \tilde{\xi}^a_{\beta\alpha}(\textbf{k}) \equiv i(\beta\textbf{k}|\partial_a \alpha\textbf{k}),
\end{equation}
and $\partial_a$ indicates a partial derivative with respect to the Cartesian component $a$ of \textbf{k}. One could instead view the components of the non-Abelian Berry connection as a 2x2 matrix, with rows and columns indicating the spinor component of the bra or the ket state respectively. Then equation (\ref{Non-Abelian}) is the trace over this matrix, since we implicitly sum over spinor components. The non-Abelian Berry connection associated with the set $\{|\alpha\textbf{k}\rangle\}$ is related to the non-Abelian Berry connection associated with the set $\{ | n\textbf{k} \rangle \}$ via

\begin{equation}
    \sum_{\alpha\beta} U_{m\beta}\tilde{\xi}^a_{\beta\alpha}U^\dag_{\alpha n} = \xi^a_{mn} + \mathcal{W}^a_{mn},
\end{equation}
where

\begin{equation}
    \mathcal{W}^a_{mn} = i \sum_{\alpha} (\partial_a U_{m\alpha}) U^\dag_{\alpha n}.
\end{equation}

We now turn to new terms that explicitly involve the electron spin. Just as there is a relation (\ref{BerryConnection}) between the position matrix elements of the Wannier functions and the non-Abelian Berry connection, there is a relation between the spin matrix elements in the Wannier function basis and the \textbf{k} dependent spin matrix elements in the Bloch cell-periodic basis. This relationship follows from equations (\ref{Blochfunctions}, \ref{unitcellintegral}, \ref{WannierKet}), and dividing the integral over all space into a sum of integrals over unit cells, we have

\begin{equation}
\label{Spin1}
\begin{split}
    \frac{\hbar}{2}\int d\textbf{x} W^*_{\alpha\textbf{R},\textsf{i}}(\textbf{x}) \sigma_{\textsf{i}\textsf{j}}^a W_{\beta\textbf{0},\textsf{j}}(\textbf{x}) 
    =\frac{\Omega_{uc}}{(2\pi)^3} \int_{BZ} \hspace{-5pt} d\textbf{k} e^{i\textbf{k}\cdot\textbf{R}} \tilde{S}^a_{\alpha\beta}(\textbf{k}),
\end{split}
\end{equation}
where 
\begin{equation}
\label{Spin2}
    \tilde{S}^a_{\alpha\beta}(\textbf{k}) \equiv \frac{\hbar}{2} (\alpha\textbf{k}|\sigma^a \beta\textbf{k}).
\end{equation}
The spin matrix elements in the basis of the cell-periodic Bloch functions and the new Wannier cell-periodic functions are related by the unitary transformation

\begin{equation}
\label{Spin3}
    S^a_{mn}(\textbf{k}) = \sum_{\alpha\beta} U_{m\alpha} \tilde{S}^a_{\alpha\beta}(\textbf{k}) U^\dag_{\beta n}.
\end{equation}

Other terms explicitly involving the electron spin are the matrix elements of the velocity operator,
\begin{equation}
\label{RealSpaceVel}
\begin{split}
    \int \psi^\dag_{n'\textbf{k}'}(\textbf{x}) \left( \mathfrak{p}^a(\textbf{x}) + \epsilon^{abc}\frac{\hbar}{4mc^2} \sigma^b \frac{\partial V(\textbf{x})}{\partial x^c} \right) \psi_{n\textbf{k}}(\textbf{x}) d\textbf{x}
    \\
    = m v^a_{n'n}(\textbf{k}) \delta(\textbf{k}-\textbf{k}').
\end{split}
\end{equation}
However, as we confirm in Appendix \ref{AppendixA}, the velocity matrix elements satisfy the same relation that holds for velocity matrix elements if spin is not included \cite{Perry_Sipe}, 

\begin{equation}
\label{kspaceVel}
    v^a_{n'n}(\textbf{k}) =  \frac{\delta_{n'n}}{\hbar} \partial_a E_{n\textbf{k}} + \frac{i}{\hbar} (E_{n'\textbf{k}}-E_{n\textbf{k}}) \xi^a_{n'n}(\textbf{k}),
\end{equation}
where of course in the spin-full case the Bloch functions used in calculating all the terms in (\ref{kspaceVel}) include the effects of spin-orbit coupling. We refer to that fact that (\ref{kspaceVel}) holds whether the calculation is made for spin-full or spin-less electrons as the ``velocity matrix equvalence" (VME).  Because it holds, many of the contributions to the optical response tensors, when written in the Bloch state basis, take the same form in the spin-full calculation as in the spin-less calculation. We will identify these equivalences as they arise below. 

If a vector potential is applied, Wannier functions multiplied by a generalized Peierls phase factor arise, and these modified Wannier functions are in general neither orthonormal nor gauge invariant in the electromagnetic sense.
We use Lowdin's method of symmetric orthogonalization \cite{Lowdin} to construct an orthonormal set of functions \{$\bar{W}_{\alpha\textbf{R}}(\textbf{x},t)$\},
the ``adjusted Wannier functions". 
These can be written as
\begin{equation}
\label{adjustedWannierFunctions}
    \bar{W}_{\alpha\textbf{R}}(\textbf{x},t) = e^{i\Phi(\textbf{x},\textbf{R};t)} \chi_{\alpha\textbf{R}}(\textbf{x},t),
\end{equation}
where the set of functions \{$\chi_{\alpha\textbf{R}}$\} are gauge-invariant in the electromagnetic sense, and the phase factor involves the generalized Peierls phase 
$\Phi(\textbf{x},\textbf{R};t)$, a line integral of the vector potential over an arbitrary path from $\textbf{R}$ to $\textbf{x}$, 

\begin{equation}
    \Phi(\textbf{x},\textbf{R};t) \equiv \frac{e}{\hbar c} \int s^i(\textbf{w};\textbf{x},\textbf{R}) A^i(\textbf{w},t) d\textbf{w},
\end{equation}
where the function \textbf{s}(\textbf{w};\textbf{x},\textbf{R}) is a  so-called ``relator" \cite{Perry_Sipe}, and is defined as

\begin{equation}
\label{relator-s}
    s^i(\textbf{w};\textbf{x},\textbf{R}) = \int_{C(\textbf{x},\textbf{R})} dz^i \delta(\textbf{w}-\textbf{z}),
\end{equation}
where $C(\textbf{x},\textbf{R})$ specifies a path from \textbf{R} to \textbf{x}. In a perturbation expansion for the functions \{$\chi_{\alpha\textbf{R}}$\}, in powers of the magnetic field, the first two terms are given by

\begin{equation}
\begin{split}
    &\chi_{\alpha\textbf{R},\textsf{i}}(\textbf{x},t) = W_{\alpha\textbf{R},\textsf{i}}(\textbf{x})
    - \frac{i}{2} \sum_{\beta\textbf{R}',\textsf{j}} W_{\beta\textbf{R}',\textsf{i}} 
    \\
    &\times
    \int W^*_{\beta\textbf{R}',\textsf{j}}(\textbf{z}) \Delta(\textbf{R}',\textbf{z},\textbf{R};t) W_{\alpha\textbf{R},\textsf{j}} (\textbf{z})+ ...
\end{split}
\end{equation}
Here the function $\Delta(\textbf{x},\textbf{y},\textbf{z};t)$ is $\Phi(\textbf{z},\textbf{x};t)+\Phi(\textbf{x},\textbf{y};t)+\Phi(\textbf{y},\textbf{z};t)$, which is simply a closed line integral of the vector potential. By Stokes' theorem this is an integral of the magnetic flux passing through the surface identified by the paths connecting \textbf{x}, \textbf{y}, and \textbf{z}. Both $\Phi(\textbf{x},\textbf{R};t)$ and $\Delta(\textbf{x},\textbf{y},\textbf{z})$ have been discussed earlier \cite{Perry_Sipe}. Indeed, the strategy here follows what has been done there for spinless particles, with the change that quantities such as $\chi_{\alpha\textbf{R},\textsf{i}}(\textbf{x},t)$ and $W_{\alpha\textbf{R},\textsf{i}}(\textbf{x})$ carry spinor indices.

\subsection{Global Green function}

We now turn to establishing Green function expressions in a form that will allow us to focus on individual lattice site quantities.  We begin by expanding our field operators in the Heisenberg picture in a basis formed by the ``adjusted Wannier functions" (\ref{adjustedWannierFunctions}), 

\begin{equation}
    \hat{\psi}(\textbf{x},t) = \sum_{\alpha,\textbf{R}} \hat{a}_{\alpha\textbf{R}}(t) \begin{bmatrix}
    \bar{W}_{\alpha\textbf{R},\uparrow}(\textbf{x},t)
    \\
    \bar{W}_{\alpha\textbf{R},\downarrow}(\textbf{x},t)
    \end{bmatrix}
    = \begin{bmatrix}
    \hat{\psi}_\uparrow(\textbf{x},t)
    \\
    \hat{\psi}_\downarrow(\textbf{x},t)
    \end{bmatrix},
\end{equation}
where from the anticommutation relations that the field operators satisfy it follows that
\begin{equation}
\begin{split}
    \{\hat{a}_{\alpha\textbf{R}}(t),\hat{a}_{\beta\textbf{R}'}(t)\} &= 0,
    \\
    \{ \hat{a}_{\alpha\textbf{R}}(t),\hat{a}^\dag_{\beta\textbf{R}'}(t)\} &= \delta_{\alpha\beta} \delta_{\textbf{R}\textbf{R}'}.
\end{split}
\end{equation}
The lesser, equal time single-particle Green function matrix then has the spin-components

\begin{equation}
\label{MatrixGreen}
    G^{mc}_\textsf{ij}(\textbf{x},\textbf{y};t) = i \langle \hat{\psi}^\dag_\textsf{j}(\textbf{y},t) \hat{\psi}_\textsf{i}(\textbf{x},t) \rangle.
\end{equation}
Here ``mc" denotes ``minimal coupling," and the dynamics of the field operators are governed by the Hamiltonian (\ref{scriptH}). To move to a Green function matrix that is gauge-invariant with respect to the electromagnetic potentials, we introduce a global Green function according to 

\begin{equation}
    G^{gl}(\textbf{x},\textbf{y};t) = e^{-i\Phi(\textbf{x},\textbf{y};t)}G^{mc}(\textbf{x},\textbf{y};t).
\end{equation}
\begin{widetext}
 Following manipulations used in the spinless problem \cite{Perry_Sipe}, the dynamics of $G^{gl}$ are found to be governed by

\begin{equation}
\begin{split}
    i\hbar \frac{\partial G^{gl}_\textsf{ij}(\textbf{x},\textbf{y};t)}{\partial t} = \mathcal{K}_\textsf{ik}(\textbf{x},\textbf{y};t) G^{gl}_\textsf{kj}(\textbf{x},\textbf{y};t) -
    G^{gl}_\textsf{ik}(\textbf{x},\textbf{y};t)\overleftarrow{\mathcal{K}}_\textsf{kj}(\textbf{y},\textbf{x};t) - e\Omega^0_{\textbf{y}}(\textbf{x},t) G^{gl}_\textsf{ij}(\textbf{x},\textbf{y};t),
\end{split}
\end{equation} 
where the modified Hamiltonian differential operator matrix elements are

\begin{equation}
\begin{split}
    \mathcal{K}_\textsf{ij}(\textbf{x},\textbf{y},t) = \frac{(\boldsymbol{\mathfrak{p}}(\textbf{x})-\frac{e}{c}\boldsymbol\Omega_\textbf{y}(\textbf{x},t))^2}{2 m} \delta_\textsf{ij} 
    +  \text{V}(\textbf{x})\delta_\textsf{ij}
    - \frac{e\hbar}{2mc} \boldsymbol\sigma_\textsf{ij}\cdot \textbf{B}(\textbf{x},t) - \frac{e\hbar}{2mc} \boldsymbol\sigma_\textsf{ij}\cdot \textbf{B}_\text{static}(\textbf{x})
    \\
    + \frac{\hbar}{4m^2c^2} \boldsymbol\sigma_\textsf{ij} \cdot \left(\nabla \text{V}(\textbf{x})\times (\boldsymbol{\mathfrak{p}}(\textbf{x}) - \frac{e}{c}\boldsymbol\Omega_\textbf{y}(\textbf{x},t)) \right),
\end{split}
\end{equation}
\end{widetext}
and where the matrix $\overleftarrow{\mathcal{K}}(\textbf{x},t)$ is identical to $\mathcal{K}(\textbf{x},t)$, except that in the matrix elements of $\overleftarrow{\mathcal{K}}_\textsf{ij}(\textbf{x},t)$ the momentum operator is complex conjugated and taken as acting to the left. Where once there was the applied vector potential accompanying the momentum, there is now the new quantity $\boldsymbol\Omega_{\textbf{y}}(\textbf{x},t)$, and the applied scalar potential term has been replaced by one involving $\Omega_\textbf{y}^0(\textbf{x},t)$. These quantities depend only on the electromagnetic fields and not the potentials, and are given by

\begin{equation}
    \Omega^0_\textbf{y}(\textbf{x},t) \equiv \int s^i(\textbf{w};\textbf{x},\textbf{y}) E^i(\textbf{w},t) d\textbf{w},
\end{equation}
\begin{equation}
    \Omega_\textbf{y}^k(\textbf{x},t) \equiv \int \alpha^{lk}(\textbf{w};\textbf{x},\textbf{y}) B^l(\textbf{w},t) d\textbf{w},
\end{equation}
where we have employed another ``relator" $\alpha^{jk}(\textbf{w};\textbf{x},\textbf{y})$ \cite{HealyQuantum}, defined as
\begin{equation}
\label{relator-alpha}
    \alpha^{jk}(\textbf{w};\textbf{x},\textbf{y}) = \epsilon^{jmn} \int_{C(\textbf{x},\textbf{y})} dz^m \frac{\partial z^n}{\partial x^k} \delta(\textbf{w}-\textbf{z}).
\end{equation}
The quantities $\boldsymbol\Omega_\textbf{y}(\textbf{x},t)$, which are dependent on the magnetic field, and $\Omega_\textbf{y}^0(\textbf{x},t)$, which are dependent on the electric field, have been discussed earlier \cite{Perry_Sipe}.  In moving from $G^{mc}_\textsf{ij}(\textbf{x},\textbf{y};t)$ to $G^{gl}_\textsf{ij}(\textbf{x},\textbf{y};t)$, the gauge freedom of the electromagnetic potentials has been replaced by a freedom in choosing the path C(\textbf{x},\textbf{y}) involved in the definitions of the relators \cite{PZWNewPaper}.
 
\subsection{Site quantities}

To move to the introduction of site quantities, we begin with expressions for the full charge and current densities,

\begin{equation}
\begin{split}
    \langle \hat{\rho} (\textbf{x},t) \rangle 
    = -ie[G_\textsf{ii}^{gl}(\textbf{x},\textbf{y};t)]_{\textbf{y}\rightarrow \textbf{x}}+\rho^\text{ion}(\textbf{x}),
\end{split}
\end{equation}
and
\begin{equation}
\begin{split}
    \langle \hat{\textbf{j}}(\textbf{x},t) \rangle 
    = -ie\left[ \mathcal{J}_\textsf{ij}^{gl}(\textbf{x},\textbf{y};t) G_\textsf{ji}^{gl}(\textbf{x},\textbf{y};t) \right]_{\textbf{y}\rightarrow \textbf{x}}.
\end{split}
\end{equation}
Here we have introduced a differential operator for the global charge current,

\begin{equation}
\begin{split}
    \mathcal{J}^{gl}_\textsf{ij}(\textbf{x},\textbf{y};t) 
    =\textbf{J}_\textsf{ij}^{gl}(\textbf{x},\boldsymbol{\mathfrak{p}}(\textbf{x},\textbf{y};t)) + \textbf{J}_\textsf{ij}^{gl}(\textbf{y},\boldsymbol{\mathfrak{p}}^*(\textbf{y},\textbf{x};t)),
\end{split}
\end{equation}
with
\begin{equation}
\begin{split}
    \textbf{J}^{gl}_\textsf{ij}(\textbf{x},\boldsymbol{\mathfrak{p}}(\textbf{x},\textbf{y};t)) = \frac{e}{2m} \mathfrak{p}(\textbf{x},\textbf{y};t)\delta_\textsf{ij} 
    + \hat{e}^c \epsilon_{abc} \frac{e\hbar}{2m} \sigma^b_\textsf{ij} \frac{\partial }{\partial x_a}
    \\
    + \hat{e}^c \epsilon_{abc} \frac{e \hbar}{8m^2c^2} \sigma_\textsf{ij}^a (\nabla V(\textbf{x}))^b.
\end{split}
\end{equation}
and where we have defined a modified momentum operator by

\begin{equation}
    \mathfrak{p}(\textbf{x},\textbf{y};t) = \mathfrak{p}(\textbf{x}) - \frac{e}{c}\boldsymbol\Omega_\textbf{y}(\textbf{x},t).
\end{equation}

Expanding the minimal coupling Green function in the adjusted Wannier function basis, we have

\begin{equation}
\begin{split}
    G^{gl}_\textsf{ij}(\textbf{x},\textbf{y};t) = i e^{-i\Phi(\textbf{x},\textbf{y};t)} \sum_{\alpha, \beta, \textbf{R}, \textbf{R}'} \langle \hat{a}^\dag_{\beta \textbf{R}'}(t) \hat{a}_{\alpha \textbf{R}}(t) \rangle 
    \\
    \times \bar{W}^*_{\beta \textbf{R}';\textsf{j}}(\textbf{y},t) \bar{W}_{\alpha \textbf{R} ; \textsf{i}}(\textbf{x},t),
\end{split}
\end{equation}
and we define the components of a single particle density matrix as

\begin{equation}
    \eta_{\alpha\textbf{R};\beta\textbf{R}'}(t) \equiv \langle \hat{a}^\dag_{\beta\textbf{R}'}(t) \hat{a}_{\alpha\textbf{R}}(t)\rangle e^{i\Phi(\textbf{R}',\textbf{R};t)}.
\end{equation}
with the phase
factor chosen to ensure that the 
dynamics of the single particle density matrix is gauge-invariant in the electromagnetic sense, as was done in the introduction of 
the global Green function. The dynamical equations for the $\eta_{\alpha\textbf{R};\beta\textbf{R}'}(t)$ were derived earlier \cite{Perry_Sipe} for systems where the spin degree of freedom is neglected. Here the derivation follows along those lines, but with the use of $\mathcal{K}(\textbf{x},t)$ as the differential operator for the Hamiltonian, and the use of spinor states in the construction of the Green function. For completeness we give the result in Appendix \ref{AppendixB}.                      

We can then decompose the global Green function into site specific Green functions labelled by a lattice site \textbf{R}, 

\begin{equation}
    G^{gl}_\textsf{ij}(\textbf{x},\textbf{y};t) = \sum_\textbf{R} e^{-i\Delta(\textbf{x},\textbf{y},\textbf{R};t)}G^\textbf{R}_\textsf{ij}(\textbf{x},\textbf{y};t),
\end{equation}
where

\begin{equation}
\begin{split}
    G^\textbf{R}_\textsf{ij}(\textbf{x},\textbf{y};t)
    =
    \frac{i}{2} &\sum_{\alpha,\beta,\textbf{R}'} \eta_{\alpha \textbf{R};\beta\textbf{R}'}(t) e^{i\Delta(\textbf{R}',\textbf{y},\textbf{R};t)}
    \\
    &\times \chi^*_{\beta\textbf{R}';\textsf{j}}(\textbf{y},t)\chi_{\alpha\textbf{R};\textsf{i}}(\textbf{x},t) 
    \\
    +\frac{i}{2}&\sum_{\alpha,\beta,\textbf{R}'} \eta_{\beta\textbf{R}';\alpha \textbf{R}}(t)e^{i\Delta(\textbf{R},\textbf{x},\textbf{R}';t)}
    \\
    &\times \chi^*_{\alpha \textbf{R};\textsf{j}}(\textbf{y},t) \chi_{\beta\textbf{R}';\textsf{i}}(\textbf{x},t),
\end{split}
\end{equation}

and we can then write

\begin{equation}
    \begin{split}
        \langle \hat{\rho}(\textbf{x},t) \rangle &= \sum_\textbf{R} \rho_\textbf{R}(\textbf{x},t), \\
        \langle \hat{\textbf{j}}(\textbf{x},t)\rangle &= \sum_\textbf{R} \textbf{j}_\textbf{R}(\textbf{x},t).
    \end{split}
\end{equation}
Site quantities such as $\textbf{j}_\textbf{R}(\textbf{x},t)$ in general involve differential operators acting on the site Green functions $G^\textbf{R}(\textbf{x},\textbf{y};t)$. Expressions for the site charge and current density are found in Appendix \ref{AppendixB}. Again, their derivation largely follows what was done earlier for spinless systems \cite{Perry_Sipe}.

And again as in the spinless case, we must also identify a free charge density $\rho_F(\textbf{x},t)$ and a free charge current density $\textbf{j}_F(\textbf{x},t)$ \cite{Perry_Sipe}. ``Free" charges at each site are constructed by placing all the charge associated with a site \textbf{R} at that site. Since the total charge associated with each site is not necessarily constant in time, link currents are identified, and from these we identify the free current density.  The expressions for these quantities follow those for spinless systems, and can be found in Appendix \ref{AppendixB}. 

The charge density $\rho_\textbf{R}(\textbf{x},t)$ associated with each site can be considered as the sum of the total charge associated with site \textbf{R} localized at that lattice site (\ref{esitecharg}), and contributions from a microscopic polarization field associated with the site that captures the multipole moments of the charge distribution, 

\begin{equation}\label{MicroPolarization}
    \textbf{p}_\textbf{R}(\textbf{x},t) \equiv \int \textbf{s}(\textbf{x};\textbf{y},\textbf{R}) \rho_\textbf{R}(\textbf{y},t)  d\textbf{y}.
\end{equation}
We also introduce magnetization fields associated with each site, which are further split into three contributions,

\begin{equation}
\label{SiteMagnetizations}
    \textbf{m}_\textbf{R}(\textbf{x},t) \equiv \bar{\textbf{m}}_\textbf{R}(\textbf{x},t) + \tilde{\textbf{m}}_\textbf{R}(\textbf{x},t) + \breve{\textbf{m}}_\textbf{R}(\textbf{x},t).
\end{equation}
These are the ``atomic," ``itinerant," and ``spin" contributions respectively. The atomic magnetization is related to the site current density $\textbf{j}_\textbf{R}(\textbf{x},t)$ in the way that the magnetization in an isolated atom would be related to its current density. The itinerant magnetization arises because there are corrections to this in a solid, since the sites are not isolated \cite{Magnetization_PI,Magnetization_MbI}; this is discussed in Appendix \ref{AppendixB}. For brevity we omit these expressions in the main body of the text, since their derivation follows what was done earlier \cite{Perry_Sipe}, with the understanding that operators such as the charge current and the Hamiltonian matrix elements now involve spin-orbit coupling effects, and the Wannier functions are now spinors. The spin contribution to the magnetization, however, is a new term here and arises due to including the spin degree of freedom. It is given by
\begin{equation}
\label{SpinMag}
\begin{split}
    \breve{\textbf{m}}_\textbf{R}(\textbf{x},t) = \frac{e\hbar}{4mc} \sum_{\alpha,\beta,\textbf{R}',\textbf{R}''} \left(\delta_{\textbf{R}\textbf{R}'} + \delta_{\textbf{R}\textbf{R}''} \right) e^{i\Delta(\textbf{R}',\textbf{x},\textbf{R}'')}
    \\
    \times \chi^*_{\beta\textbf{R}';\textsf{j}}(\textbf{x},t) \boldsymbol\sigma_\textsf{ji} \chi_{\alpha\textbf{R}'';\textsf{i}}(\textbf{x},t) \eta_{\alpha\textbf{R}'';\beta\textbf{R}'}(t).
\end{split}
\end{equation}
The total microscopic polarization and magnetization fields are simply the sum over their respective site quantities,

\begin{equation}
\label{TotalPolandMag}
\begin{split}
    \textbf{p}(\textbf{x},t) &= \sum_\textbf{R} \textbf{p}_\textbf{R}(\textbf{x},t),
    \\
    \textbf{m}(\textbf{x},t) &= \sum_\textbf{R} \textbf{m}_\textbf{R}(\textbf{x},t).
\end{split}
\end{equation}

We can write the expectation value of the total charge density and current density in terms of the polarization, magnetization and free charge and current:

\begin{equation}
\label{microscopic}
\begin{split}
    \langle \hat{\rho}(\textbf{x},t) \rangle  &= -\nabla \cdot \textbf{p}(\textbf{x},t) + \rho_F(\textbf{x},t),
    \\
    \langle \hat{\textbf{j}}(\textbf{x},t) \rangle &= \frac{\partial \textbf{p}(\textbf{x},t)}{\partial t} + c\nabla \times \textbf{m}(\textbf{x},t) + \textbf{j}_F(\textbf{x},t).
\end{split}
\end{equation}
The expressions for all the aforementioned site quantities, and for the free charge and current densities, can be found in Appendix \ref{AppendixB}. 

\section{Multipole Expansion}\label{MultipoleExpansion}

At this point, spatial averaging can be employed to construct the macroscopic version of the expressions (\ref{microscopic}) for the microscopic charge and current densities; the treatment of light propagating through the crystal, taking into account its variation over a unit cell if necessary, can then be addressed \cite{PerryOptical}. 

To treat the linear response including spatially varying fields the microscopic polarization is expanded to include the quadrupole moment
\begin{equation}
    p^{i}_\textbf{R}(\textbf{x},t) = \mu^i_\textbf{R}(t) \delta(\textbf{x}-\textbf{R}) - q^{ij}_\textbf{R}(t) \frac{\partial \delta(\textbf{x}-\textbf{R})}{\partial x^j} +...,
\end{equation}
where the electric dipole moment is
\begin{equation}
    \mu^i_\textbf{R}(t) \equiv \int d\textbf{y} (y^i-R^i) \rho_\textbf{R}(\textbf{y},t) ,
\end{equation}
and the electric quadrupole moment is
\begin{equation}
    q^{ij}_\textbf{R}(t) \equiv \frac{1}{2} \int d\textbf{y} (y^i-R^i)(y^j-R^j) \rho_\textbf{R}(\textbf{y},t),
\end{equation}
each associated with lattice site \textbf{R}. This is accomplished by expanding the relator $\textbf{s}(\textbf{x};\textbf{y},\textbf{R})$ appearing in the equation (\ref{MicroPolarization}) for the site polarization. For the moment we only consider the magnetic dipole moment and neglect higher order moments, so the microscopic magnetization fields are written as
\begin{equation}
    m^i_\textbf{R}(\textbf{x},t) = \nu^i_\textbf{R}(t) \delta(\textbf{x}-\textbf{R}) + ...,
\end{equation}
where the magnetic dipole moment is
\begin{equation}
\label{sitemagdipole}
    \nu^i_\textbf{R}(t) \equiv \int d\textbf{x} \Big(\bar{m}^i_\textbf{R}(\textbf{x},t) + \tilde{m}^i_\textbf{R}(\textbf{x},t) + \breve{m}^i_\textbf{R}(\textbf{x},t) \Big)
\end{equation}

Before any fields are applied we can obtain expressions for the ground state dipole moments. If inversion symmetry is broken there is the possibility for a spontaneous polarization, and if time-reversal symmetry is broken before the application of any fields, a spontaneous magnetization is possible. 

The expression for the ground state polarization (see, e.g., Mahon and Sipe \cite{PerryOptical}) is unchanged upon including spin-orbit coupling in this formalism, although of course spinor Wannier functions that capture the effects of spin-orbit coupling must be used in its calculation. However, beyond the effects of spin-orbit coupling the magnetization gains an explicitly spin-dependent contribution, and so we turn now to the contributions to the magnetization. 

The total magnetic dipole moment of a site is the sum of the atomic, itinerant, and spin contributions (\ref{SiteMagnetizations}). We evaluate the ground state expressions by setting the applied fields to zero in (\ref{sitemagdipole}). The real space-expressions for the ``atomic" and ``itinerant" ground state magnetization are different from the previous treatment \cite{Perry_Sipe} of spinless systems, because the Wannier functions are now spinors, and because there are spin-orbit contributions to the velocity operator. 
The expression for the atomic magnetization that follows from the atomic contribution to the magnetic dipole moment is
\begin{equation}
\label{nubar}
\begin{split}
    &\bar{M}^{i(0)} = \frac{\epsilon^{iab}}{\Omega_{uc}} \frac{1}{c} \sum_{\alpha} f_{\alpha} \int d\textbf{x} (x^a-R^a) W^*_{\alpha\textbf{R}}(\textbf{x})
    \\
    &\times \left(\frac{e}{2m}\mathfrak{p}^b(\textbf{x}) + \epsilon^{blm} \frac{e\hbar}{8m^2c^2} \sigma^l \nabla_m \text{V}(\textbf{x}) \right) W_{\alpha\textbf{R}}(\textbf{x}),
\end{split}
\end{equation}
and the corresponding expression for the itinerant magnetization is

\begin{equation}
\label{nutilde}
\begin{split}
    \tilde{M}^{i(0)} = 
    \frac{e}{2\hbar c} \frac{\epsilon^{iab}}{\Omega_{uc}} \sum_{\alpha\mu\textbf{R}'} f_{\alpha} \text{Im} \Bigg\{ 
    \int d\textbf{x} R'^a x^b
    \\
    \times W^*_{\mu\textbf{R}'}(\textbf{x})
    W_{\alpha\textbf{0}}(\textbf{x})
    H^{(0)}_{\alpha\textbf{0};\mu\textbf{R}'}
    \Bigg\},
\end{split}
\end{equation}
where the zero-order Hamiltonian matrix elements are

\begin{equation}
    H^{(0)}_{\alpha\textbf{R};\beta\textbf{R}'} = \int W^*_{\alpha\textbf{R}}(\textbf{x}) \mathcal{H}^0(\textbf{x}) W_{\beta\textbf{R}'}(\textbf{x}) d\textbf{x}.
\end{equation}
These expressions can then be converted to \textbf{k}-space integrals \cite{Perry_Sipe}. Despite the appearance of the spin-orbit contribution in (\ref{nubar}), the \textbf{k}-space expression for ${\bar{\textbf{M}}^{(0)}}$ is formally the same as if there were no spin, because of VME.

The contribution to the site magnetic dipole moment due to the spin degree of freedom is just as one would expect,

\begin{equation}
\begin{split}
    \breve{\nu}^{(0)}_\textbf{R} = \frac{e\hbar}{2mc} \sum_{\alpha} f_{\alpha} \int d\textbf{x} W^*_{\alpha\textbf{R}}(\textbf{x}) \boldsymbol\sigma W_{\alpha\textbf{R}}(\textbf{x}),
\end{split}
\end{equation}
and using equations (\ref{Spin1}) and (\ref{Spin3})  we find the spin contribution to the magnetization is
\begin{equation}
    \breve{\textbf{M}}^{(0)} = \frac{e}{ mc}  \sum_n f_n \int \frac{d\textbf{k}}{(2\pi)^3} \textbf{S}_{nn}(\textbf{k}),
\end{equation}
which is gauge-invariant in the  sense of how one constructs the Wannier functions.  

\section{Linear Response}\label{LinearResponse}

To consider the linear response we begin by expanding time dependent quantities in a Fourier series
\begin{equation}
    g(t) = \sum_\omega e^{-i\omega t}g(\omega),
\end{equation}
and find that the relevant terms to the linear response are

\begin{equation}
\label{multipoleresponses}
\begin{split}
    \mu^i_\textbf{R}(t) &= \mu_\textbf{R}^{i(0)} + \Omega_{uc} \sum_{\omega} e^{-i\omega t} \Big( \chi^{il}_E(\omega) E^l(\textbf{R},\omega) 
    \\
    &+ \gamma^{ijl}(\omega) F^{jl}(\textbf{R},\omega) + \alpha^{il}_{\mathcal{P}}(\omega) B^l(\textbf{R},\omega)+...\Big),
    \\
    q^{ij}_\textbf{R}(t) &= q^{ij(0)}_\textbf{R} + \Omega_{uc} \sum_{\omega} e^{-i\omega t} \chi^{ijl}_\mathcal{Q}(\omega) E^l(\textbf{R},\omega) + ... 
    \\
    \nu_\textbf{R}^i(t) &= \nu_\textbf{R}^{i(0)} + \Omega_{uc} \sum_{\omega} e^{-i\omega t} \alpha^{li}_\mathcal{M}(\omega) E^l(\textbf{R},\omega) + ...
\end{split}
\end{equation}
where 
\begin{equation}
    F^{jl}(\textbf{x},\omega) = \frac{1}{2}\Big(\frac{\partial E^j(\textbf{x},\omega)}{\partial x^l} + \frac{\partial E^l(\textbf{x},\omega)}{\partial x^j}\Big)
\end{equation}
is the symmetric derivative of the electric field.

Implementing the macroscopic averaging technique discussed in Mahon et al. \cite{PerryOptical} the macroscopic polarization and magnetization can be written as

\begin{equation}
\begin{split}
    P^i(\textbf{x},t) &= \frac{\mu_\textbf{R}^{i(0)}}{\Omega_{uc}}  + \sum_{\omega} e^{-i\omega t}\Big( \chi^{il}_E(\omega) E^l(\textbf{x},\omega) 
    \\
    &+ \gamma^{ijl}(\omega) F^{jl}(\textbf{x},\omega) + \beta^{il}_\mathcal{P}(\omega) B^l(\textbf{x},\omega)
    \\
    &-\chi^{ijl}_Q(\omega) \frac{\partial E^l(\textbf{x},\omega)}{\partial x^j} + ...\Big)
    \\
    M^i(\textbf{x},t) &= \frac{\nu_\textbf{R}^{i(0)}}{\Omega_{uc}} + \sum_{\omega} e^{-i\omega t}\alpha^{li}_{\mathcal{M}}(\omega) E^l(\textbf{x},\omega) + ...
\end{split}
\end{equation}
\vspace{7pt}
These expressions for the macroscopic polarization and magnetization can then be implemented in the macroscopic version of (\ref{microscopic}) to obtain

\begin{equation}
\label{OpticalConductivityCurrent}
\begin{split}
    J^i(\textbf{x},\omega) &= -i\omega P^i(\textbf{x},\omega) + c \epsilon^{ijk} \frac{\partial M^k(\textbf{x},\omega)}{\partial x^j}
    \\
    &= \sigma^{il}(\omega) E^l(\textbf{x},\omega) -i\sigma^{ilj}(\omega) \frac{\partial E^l(\textbf{x},\omega)}{\partial x^j} + ...
\end{split}
\end{equation}
In going to the second line the response of the polarization to a magnetic field can be included in the tensor $\sigma^{ilj}(\omega)$ due to Faraday's law. The conductivity tensors can then be expressed as

\begin{equation}
\begin{split}
    \sigma^{il}(\omega) &= -i\omega \chi^{il}_E(\omega)
    \\
    \sigma^{ilj}(\omega) &= \omega\gamma^{ijl}(\omega) -ic\alpha^{ia}_\mathcal{P}(\omega) \epsilon^{ajl} -\omega \chi^{ijl}_{Q}(\omega)
    \\
    &+ic\epsilon^{ija}\alpha^{la}_{\mathcal{M}}(\omega),
\end{split}
\label{response}
\end{equation}
where a detailed discussion of the various contributions to the optical conductivity tensor are discussed in an earlier work \cite{PerryOptical}. In that earlier work the $\alpha^{il}_{\mathcal{P}/\mathcal{M}}(\omega)$ only included terms that vanish as the frequency vanishes; the Chern-Simons and cross-gap contributions, which survive at all frequencies, were tabulated separately. Here we include them in $\alpha^{il}_{\mathcal{P}/\mathcal{M}}(\omega)$, so that the expressions (\ref{FiniteFreqMP}) capture the full response of $\textbf{P}(\omega)$  to $\textbf{B}(\omega)$ and $\textbf{M}(\omega)$ to $\textbf{E}(\omega)$.
In section \ref{OpticalConductivitySection} we give the full expressions for $\sigma^{il}(\omega)$ and $\sigma^{ilj}(\omega)$, including the spin contribution.

\vspace{10pt}
\subsection{First order response}

Linear response of the polarization, magnetization, or quadrupolarization can arise because the single particle density matrix responds to the external electromagnetic field, and also because the matrix elements involved in the quantity being calculated can depend on the electromagnetic field. We refer to these contributions as ``dynamical" and ``compositional" respectively \cite{Perry_Sipe}, and label the contributions to a polarization, magnetization, or quadrupolarization by $I$ and $II$ respectively. We begin by collecting the responses of the single particle density matrix to the electromagnetic field,
using $(E)$, $(B)$ and $(F)$ to indicate a response to the electric, magnetic or symmetrized derivatives of the electric field respectively.

The first order response of the single particle density matrix to an applied electric field at arbitrary frequency 
\begin{widetext}
\noindent is
\begin{equation}
\begin{split}
    \eta^{(E)}_{\alpha\textbf{R};\beta\textbf{R}'}(\omega) = eE^l(\textbf{R}_a;\omega) \Omega_{uc} \sum_{mn} f_{nm} \int_{BZ}  \frac{d\textbf{k}}{(2\pi)^3} 
    \frac{ e^{i\textbf{k}\cdot(\textbf{R}-\textbf{R}')} U^\dag_{\alpha m} \xi^l_{mn} U_{n\beta} }{\Delta_{mn}(\textbf{k})-\hbar(\omega+i0^+)},
\end{split}
\end{equation}
where $f_{nm} = f_n - f_m$, and $\Delta_{mn}(\textbf{k}) = E_{m\textbf{k}}-E_{n\textbf{k}}$. The electromagnetic fields and their derivatives are evaluated at an arbitrary lattice site $\textbf{R}_a$, which when implemented in the expressions for site quantities associated with a lattice site $\textbf{R}$ provides a natural choice $\textbf{R}_a = \textbf{R}$. 

The first order response of the single particle density matrix to an arbitrary frequency magnetic field is given by the expression found in Mahon et al. \cite{PerryOptical} and an added spin dependent contribution.  The total result is

\begin{equation}
\label{etatoB}
\begin{split}
    \eta^{(B)}_{\alpha\textbf{R}'';\beta\textbf{R}'}(\omega) =& \frac{e\Omega_{uc}}{4\hbar c} \epsilon^{lab} B^l(\textbf{R}_a,\omega) \sum_{mn} f_{nm} \int_{BZ} \frac{d\textbf{k}}{(2\pi)^3} e^{i\textbf{k}\cdot(\textbf{R}''-\textbf{R}')} U^\dag_{\alpha m} \mathcal{B}^{ab}_{mn}(\textbf{k},\omega) U_{n\beta} 
    \\
    &+\frac{e\Omega_{uc}}{4\hbar c} \epsilon^{lab} B^l(\textbf{R}_a,\omega) \sum_{mn} f_{nm} \int_{BZ} \frac{d\textbf{k}}{(2\pi)^3} \frac{ e^{i\textbf{k}\cdot(\textbf{R}''-\textbf{R}')} \Delta_{mn}(\textbf{k}) \xi^b_{mn} }{\Delta_{mn}(\textbf{k})-\hbar(\omega+i0^+)} \Bigg[ (\partial_a U^\dag_{\alpha m} U_{n\beta} - U^\dag_{\alpha m} (\partial_a U_{n\beta}) \Bigg]
    \\
    &-\frac{i\omega e \Omega_{uc}}{4c} \epsilon^{lab} B^l(\textbf{R}_a,\omega) \sum_{mn} f_{nm} \int_{BZ} \frac{d\textbf{k}}{(2\pi)^3} \frac{e^{i\textbf{k}\cdot(\textbf{R}''-\textbf{R}')} U^\dag_{\alpha m} \xi^a_{mn} U_{n\beta}}{\Delta_{mn}(\textbf{k})-\hbar(\omega+i0^+)} \Bigg[ (R''^b-R^b_a) + (R'^b-R^b_a)\Bigg]
    \\
    &+\frac{e\Omega_{uc}}{mc} B^l(\textbf{R}_a,\omega) \sum_{mn} f_{nm} \int_{BZ} \frac{d\textbf{k}}{(2\pi)^3}  \frac{ e^{i\textbf{k}\cdot(\textbf{R}''-\textbf{R}')} U^\dag_{\alpha m} S^l_{mn} U_{n\beta} }{\Delta_{mn}(\textbf{k})  -\hbar (\omega+i0^+)},
\end{split}
\end{equation}
where
\begin{equation}
\begin{split}
    \mathcal{B}^{ab}_{mn}(\textbf{k},\omega) = i \sum_s \Bigg[ \frac{ \Delta_{sn}(\textbf{k}) }{\Delta_{mn}(\textbf{k})-\hbar(\omega+i0^+)} \xi^a_{ms}\xi^b_{sn} + \frac{\Delta_{sm}(\textbf{k})}{\Delta_{mn}(\textbf{k})-\hbar(\omega+i0^+)} \xi^a_{ms}\xi^b_{sn} \Bigg]
    \\
    -\Bigg( 2+ \frac{\hbar\omega}{\Delta_{mn}(\textbf{k})-\hbar(\omega+i0^+)} \Bigg) \frac{\partial_a(E_{m\textbf{k}}+E_{n\textbf{k}})}{\Delta_{mn}(\textbf{k})-\hbar(\omega+i0^+)} \xi^b_{mn}.
\end{split}
\end{equation}
The last line of equation (\ref{etatoB}) arises due to the added magnetic dipole energy term in the Hamiltonian. We additionally have the response of the single particle density matrix to the symmetric derivatives of the electric field, which can be written as the same BZ integral as seen previously \cite{PerryOptical},  

\begin{equation}
\begin{split}
    \eta^{(F)}_{\alpha\textbf{R}'';\beta\textbf{R}'}(\omega) &= \frac{e\Omega_{uc}}{2} F^{jl}(\textbf{R}_a,\omega) \sum_{mn} f_{nm} \int_{BZ} \frac{d\textbf{k}}{(2\pi)^3} e^{i\textbf{k}\cdot(\textbf{R}''-\textbf{R}')} U^\dag_{\alpha m} \mathcal{F}^{jl}_{mn}(\textbf{k},\omega) U_{n\beta}
    \\
    &+ \frac{ie\Omega_{uc}}{2} F^{jl}(\textbf{R}_a,\omega) \sum_{mn} f_{nm} \int_{BZ} \frac{d\textbf{k}}{(2\pi)^3} \frac{e^{i\textbf{k}\cdot(\textbf{R}''-\textbf{R}')} \xi^l_{mn}}{\Delta_{mn}(\textbf{k})-\hbar(\omega+i0^+)} \Big[ U^\dag_{\alpha m} \partial_j U_{n\beta} - \partial_j U^\dag_{\alpha m} U_{n\beta} \Big]
    \\
    &+ \frac{e \Omega_{uc}}{2} F^{jl}(\textbf{R}_a,\omega) \sum_{mn} f_{nm} \int_{BZ} \frac{d\textbf{k}}{(2\pi)^3} \frac{e^{i\textbf{k}\cdot(\textbf{R}''-\textbf{R}')} U^\dag_{\alpha m} \xi^l_{mn} U_{n\beta} }{\Delta_{mn}(\textbf{k})-\hbar(\omega+i0^+)} \Big[ (R''^j - R^j_a) + (R'^j - R_a^j) \Big],
\end{split}
\end{equation}
with
\begin{equation}
    \mathcal{F}^{jl}_{mn}(\textbf{k},\omega) = \sum_s \frac{ \xi^j_{ms}\xi^l_{sn} }{\Delta_{mn}(\textbf{k})-\hbar(\omega+i0^+)} +i\frac{ \partial_j (E_{m\textbf{k}}+E_{n\textbf{k}})}{(\Delta_{mn}(\textbf{k})-\hbar(\omega+i0^+))^2} \xi^l_{mn}.
\end{equation}

\subsubsection{Polarization response to B}

In the spinless calculation there is both a dynamical and a compositional modification of the polarization due to an applied uniform magnetic field, $\textbf{P}^{(B)}(\omega) = \textbf{P}^{(B,I)}(\omega)+\textbf{P}^{(B,II)}(\omega)$ \cite{PerryOptical}. These retain their same form here, with the Bloch functions involved in the matrix elements now understood to be spinors. In addition, there is a new spin dependent dynamical contribution,
\begin{equation}
    P^{i,(B;\boldsymbol\sigma)}(\omega) = \frac{e^2}{mc} B^l(\omega) \sum_{mn} f_{nm} \int \frac{d\textbf{k}}{(2\pi)^3} \frac{\xi^i_{nm}S^l_{mn}}{\Delta_{mn}(\textbf{k})-\hbar(\omega+i0^+)}.
\end{equation}

\subsubsection{Spin magnetization response to E}
In general we can expect atomic, itinerant, and spin contributions to the response of the magnetization to an electric field $\textbf{M}^{(E)} = \bar{\textbf{M}}^{(E)} + \tilde{\textbf{M}}^{(E)} + \breve{\textbf{M}}^{(E)}$. All three involve dynamical modifications, and there is also a compositional modification of the itinerant contribution. 
The frequency dependent spin magnetization response to an electric field is

\begin{equation}
\label{alphaS}
    \breve{M}^{i(E,I)}(\omega) = \frac{e^2}{mc} E^l(\omega,\textbf{R}) \sum_{mn} f_{nm} \int_{BZ} \frac{d\textbf{k}}{(2\pi)^3} \frac{S^i_{nm}\xi^l_{mn}}{\Delta_{mn}(\textbf{k})-\hbar(\omega+i0^+)},
\end{equation}
which is gauge-independent.

The remaining responses outlined in equation (\ref{multipoleresponses}) can be found in a previous paper \cite{PerryOptical}, where inner products are now understood as involving spinors, and the band structure and real space velocity operator are altered by spin-orbit coupling. However, when expressed as a single integral over the BZ the functional form is the same due to VME; hence we omit repeating those results here.

\end{widetext}

\section{Frequency Dependent Magnetoelectric Polarizability
}\label{FreqMP}
We consider first the limit of uniform and static applied fields. The magnetoelectric polarizability tensor is defined as
\begin{equation}
\label{MPSym}
    \alpha^{il} =\frac{\partial P^i}{\partial B^l}\Big|_{\substack{\textbf{E} =\textbf{0} \\\textbf{B}=\textbf{0}}} = \frac{\partial M^l}{\partial E^i}\Big|_{\substack{\textbf{E}=\textbf{0} \\\textbf{B}=\textbf{0}}}.
\end{equation}
The orbital contributions to the MP tensor are usually written as the sum of an isotropic Chern-Simons contribution $\alpha^{il}_{CS}$ and a ``cross gap contribution," $\alpha^{il}_G$. In section IV of Mahon et al. \cite{PerryMagneto}, where spinless electrons were considered, the splitting of the orbital contributions to the MP tensor derived there into $\alpha^{il}_{CS}$ and $\alpha^{il}_{G}$ is performed. That strategy is valid here as well, and so we can split the full MP tensor into the familiar Chern Simons contribution, the cross gap contribution, and a new explicitly spin dependent contribution $\alpha^{il}_S$, 

\begin{equation}
\label{MPTensor}
    \alpha^{il} = \alpha^{il}_{CS} + \alpha^{il}_G + \alpha^{il}_S,
\end{equation}
where the Zeeman interaction leads to the new contribution $\alpha_S$. But it does not alter the functional form of $\alpha_{CS}^{il}+\alpha_{G}^{il}$; the first two terms on the right hand side of equation (\ref{MPTensor}), when expressed as integrals over the BZ involving the non-Abelian Berry connection matrix elements and the band dispersion, take the same form for spin-full electrons as they did for spin-less electrons \cite{PerryMagneto}, due to VME. As in the spinless electron problem, the Chern-Simons contribution is gauge dependent, and can take on values that are integer multiples of $\frac{e^2}{\hbar}$; it does not lead to any net charge or current density associated with \textbf{P} and \textbf{M}, but is linked to a quantized surface Hall conductance \cite{VanderbiltOMP,OMP2}. The cross-gap contribution vanishes unless both time reversal and inversion symmetry are broken.
The new contribution $\alpha_{S}^{il}$, which appears as a dynamical modification of the polarization if the response of the polarization to the magnetic field is calculated, or as a dynamical modification to the magnetization due to the presence of $\Omega_\textbf{R}^0$ in the Hamiltonian matrix elements (\ref{HMatrix}) if the response of the magnetization to the electric field is calculated, is given by 

\begin{equation}\label{alphaS_Tensor}
    \alpha^{il}_{S} = \frac{e^2}{mc} \sum_{mn} f_{nm} \int_{BZ} \frac{d\textbf{k}}{(2\pi)^3} \frac{ \xi^i_{nm} S^l_{mn} }{ \Delta_{mn}(\textbf{k})}. 
\end{equation}
It vanishes unless both inversion symmetry and time reversal symmetry are broken; the expression is gauge-invariant so there is no ambiguity in its value.

As expected, in the static limit the expressions for $\alpha^{il}_{CS}$ and $\alpha^{il}_{G}$ are in exact agreement with those of the ``modern theory," which are based on adiabatic and thermodynamic arguments \cite{Macro_Polar,BerryVanderbilt,VanderbiltOMP,SpinMagneto_Adiabatic}. Adding spin-orbit coupling and a Zeeman-type coupling can have the effect of breaking degeneracies, opening a gap, or changing the character of the bands. These produce ``implicit changes" to the orbital magnetoelectric polarizability $\alpha^{il}_G + \alpha^{il}_{CS}$ that are hard to quantify without specifying a model and employing the resulting energy band dispersions and Bloch functions in the expressions.

Beyond the ``implicit" effects on the orbital contribution $\alpha^{il}_G$ + $\alpha^{il}_{CS}$ there is an ``explicit" contribution to the MP, $\alpha^{il}_S$, due to the inclusion of spin. The properties of the Bloch spinors when the system obeys time reversal or inversion symmetry are generally such that $\alpha^{il}_S$ must vanish, as must $\alpha^{il}_G$. There are however, additional requirements. For $\alpha^{il}_G$ to survive, the occupied valence bands must not all have the same energy, the conduction bands must not all have the same energy, and the conduction band energies must not be the negative of the valence band energies \cite{VanderbiltOMP}. Similarly, for $\alpha^{il}_S$ to survive the energies of the electrons in the ground state must not be the same for opposite spins at all k-points. 

Turning now to the responses at finite frequency, the Zeeman interaction contribution to the response of the polarization to the magnetic field and that of the magnetization to the electric field differ. They are given respectively by
\vspace{5pt}

\begin{equation}
\label{alphaSP}
    \alpha^{il}_{S;\mathcal{P}}(\omega) = \frac{e^2}{mc} \sum_{mn} f_{nm} \int_{BZ} \frac{d\textbf{k}}{(2\pi)^3} \frac{ \xi^i_{nm} S^l_{mn} }{ \Delta_{mn}(\textbf{k}) -\hbar(\omega+i0^+)}, 
\end{equation}
and
\begin{equation}
\label{alphaSM}
    \alpha^{il}_{S;\mathcal{M}}(\omega) = \frac{e^2}{mc} \sum_{mn} f_{nm} \int_{BZ} \frac{d\textbf{k}}{(2\pi)^3} \frac{ S^l_{nm}\xi^i_{mn}}{\Delta_{mn}(\textbf{k})-\hbar(\omega+i0^+)},
\end{equation}
where however $\alpha^{il}_{S;\mathcal{P}}(0) =\alpha^{il}_{S;\mathcal{M}}(0)=\alpha^{il}_{S}$, in agreement with a result found from adiabatic variation of the Hamiltonian \cite{SpinMagneto_Adiabatic}. The symmetry constraints on the frequency dependent $\alpha^{il}_{S;\mathcal{P}}(\omega)$ and $\alpha^{il}_{S;\mathcal{M}}(\omega)$ are such that in general only inversion symmetry need be broken to lead to non-vanishing components. However, if the treatment of the Hamiltonian is limited to a model of two ``scalar" copies of each Bloch function at all k - and with the restriction to an insulator such that each copy has the same filling factor -- the $\alpha^{il}_{S;\mathcal{P}/\mathcal{M}}(\omega)$ can be shown to vanish even if inversion symmetry is broken. Thus the inclusion of spin-orbit coupling, while not breaking time-reversal symmetry, necessitates a more careful spinor treatment that can lead to a nonzero $\alpha^{il}_{S;\mathcal{P}/\mathcal{M}}(\omega)$. The symmetry arguments are laid out in more detail in Appendix \ref{AppendixD}.

The orbital contributions to the generalizations of the magneto-polarizability at finite frequency are also different, depending on whether one considers the response of the polarization to the magnetic field or that of the magnetization to the electric field. They were given earlier by Mahon et al. \cite{PerryOptical} for spin-less electrons, and can be carried over here due to VME.  In all, then, the $\alpha^{il}_{\mathcal{P}}(\omega)$ and $\alpha^{li}_{\mathcal{M}}(\omega)$ of (\ref{FiniteFreqMP}) are given by

\begin{widetext}
\begin{equation}
\begin{split}
    \alpha^{li}_{\mathcal{M}}(\omega) = \alpha^{li}_{CS} + \alpha^{li}_{G} + \alpha^{li}_{S;\mathcal{M}}(\omega)
    + \frac{e^2}{4 c} \epsilon^{iab} \sum_{mn} f_{nm} \int_{BZ} \frac{d\textbf{k}}{(2\pi)^3}
    \frac{ \omega }{\Delta_{mn}(\textbf{k})-\hbar(\omega+i0^+)}\Bigg[ 
    2 \frac{\partial_b (E_{m\textbf{k}}+E_{n\textbf{k}})}{\Delta_{mn}(\textbf{k})} \xi^a_{nm}\xi^l_{mn} 
    \\
    + i\sum_{s} \frac{\Delta_{sm}(\textbf{k})}{\Delta_{mn}(\textbf{k})} \xi^a_{ns}\xi^b_{sm}\xi^l_{mn}
    + i\sum_{s} \frac{\Delta_{ns}(\textbf{k})}{\Delta_{mn}(\textbf{k})} \xi^b_{ns}\xi^a_{sm}\xi^l_{mn} - i\xi^l_{mn} \sum_{s} \Big( \mathcal{W}^a_{ns}\xi^b_{sm}+\xi^b_{ns}\mathcal{W}^a_{sm} \Big) 
    \Bigg],
\end{split}
\end{equation}
and
\begin{equation}
\begin{split}
    \alpha^{il}_\mathcal{P}(\omega) &= \alpha^{il}_{CS} + \alpha^{il}_{G} + \alpha^{il}_{S;\mathcal{P}}(\omega) 
    + \frac{e^2}{4c} \epsilon^{lab} \sum_{mn} f_{nm} \int_{BZ} \frac{d\textbf{k}}{(2\pi)^3} \frac{ \omega }{\Delta_{mn}(\textbf{k})-\hbar(\omega+i0^+)}
    \\
    &\times\Bigg[
    i\sum_{s} \frac{\Delta_{sn}(\textbf{k})}{\Delta_{mn}(\textbf{k})}\xi^a_{ms}\xi^b_{sn}\xi^i_{nm}
    + i\sum_{s} \frac{\Delta_{sm}(\textbf{k})}{\Delta_{mn}(\textbf{k})} \xi^a_{ms}\xi^b_{sn}\xi^i_{nm}
    \\
    &+ i\xi^b_{mn}\sum_{s}\Big( \xi^i_{ns}\mathcal{W}^a_{sm}+\mathcal{W}^a_{ns}\xi^i_{sm}\Big) 
    - \Bigg(3 + \frac{\hbar \omega }{\Delta_{mn}(\textbf{k})-\hbar(\omega+i0^+)} \Bigg)\frac{\partial_a (E_{m\textbf{k}}+E_{n\textbf{k}})}{\Delta_{mn}(\textbf{k})}\xi^b_{mn}\xi^i_{nm}
    \Bigg].
\end{split}
\end{equation}

\end{widetext}
Interestingly, the finite frequency generalizations of the orbital magnetoelectric polarizability tensor 
are not gauge-invariant; however when combined with other response tensors to form the optical conductivity tensor the gauge dependence is cancelled, as we see below.  

\section{Optical Conductivity tensor}\label{OpticalConductivitySection}

With $\alpha^{il}_\mathcal{P}(\omega)$ and $\alpha^{li}_{\mathcal{M}}(\omega)$ identified, from (\ref{response}) we see that to obtain $\sigma^{il}(\omega)$ and $\sigma^{ilj}(\omega)$ we require expressions for $\chi^{il}_E(\omega)$, $\gamma^{ijl}(\omega)$, and $\chi^{ijl}_{Q}(\omega)$. These last three tensors have no explicit spin contributions, and follow from earlier work \cite{PerryOptical} with the wavefunctions in the matrix elements understood to be spinors. We find

\begin{widetext}
\begin{equation}
    \sigma^{il}(\omega) = -i\omega e^2 \sum_{nm} f_{nm} \int_{BZ} \frac{d\textbf{k}}{(2\pi)^3} \frac{  \xi^i_{nm}\xi^l_{mn} }{\Delta_{mn}(\textbf{k})-\hbar(\omega+i0^+)},
\end{equation}
and
\begin{equation}
\label{effectiveqlinearconductivity}
\begin{split}
    \sigma^{ilj}(\omega) =&-ic\alpha^{ia}_G \epsilon^{ajl} + ic\epsilon^{ijb} \alpha^{lb}_G - ic\alpha^{ia}_\mathcal{S;P}(\omega)\epsilon^{ajl} + ic\epsilon^{ijb}\alpha^{lb}_{S;\mathcal{M}}(\omega) 
    \\
    +&\frac{\omega e^2}{4} \sum_{mns'} f_{nm} \int_{BZ} \frac{d\textbf{k}}{(2\pi)^3} \frac{1}{\Delta_{mn}(\textbf{k}) -\hbar(\omega+i0^+)} 
    \Bigg( \xi^i_{nm}\xi^j_{ms}\xi^l_{sn} - \xi^i_{ns}\xi^j_{sm}\xi^l_{mn}\Bigg)
    \Bigg[ 1 + \frac{\Delta_{sn}(\textbf{k}) - \Delta_{ms}(\textbf{k})}{\Delta_{mn}(\textbf{k})}\Bigg]
    \\
    +&\frac{\omega e^2}{4} \sum_{mns'} f_{nm} \int_{BZ} \frac{d\textbf{k}}{(2\pi)^3} \frac{1}{\Delta_{mn}(\textbf{k}) -\hbar(\omega+i0^+)} 
    \Bigg( \xi^i_{nm}\xi^l_{ms}\xi^j_{sn}
    -\xi^i_{sm}\xi^l_{mn}\xi^j_{ns} \Bigg)
    \Bigg[ 1 - \frac{\Delta_{sn}(\textbf{k}) - \Delta_{ms}(\textbf{k})}{\Delta_{mn}(\textbf{k})}\Bigg]
    \\
    +&\frac{i\omega e^2}{4} \sum_{mn} f_{nm} \int_{BZ} \frac{d\textbf{k}}{(2\pi)^3} \frac{ \partial_j (E_{m\textbf{k}}+E_{n\textbf{k}}) }{\Delta_{mn}(\textbf{k})-\hbar(\omega+i0^+)} \xi^i_{nm}\xi^l_{mn}\Bigg( \frac{5}{\Delta_{mn}(\textbf{k})} + \frac{\Delta_{mn}(\textbf{k}) + \hbar \omega}{ \Delta_{mn}(\textbf{k})(\Delta_{mn}(\textbf{k})-\hbar(\omega+i0^+)) }\Bigg)
    \\
    -&\frac{i\omega e^2}{4} \sum_{mn} f_{nm} \int_{BZ} \frac{d\textbf{k}}{(2\pi)^3} \frac{2}{\Delta_{mn}(\textbf{k})} \frac{ \partial_l (E_{m\textbf{k}}+E_{n\textbf{k}})}{\Delta_{mn}(\textbf{k})-\hbar(\omega+i0^+)} \xi^i_{nm}\xi^j_{mn}
    \\
    -&\frac{i\omega e^2}{4} \sum_{mn} f_{nm} \int_{BZ} \frac{d\textbf{k}}{(2\pi)^3} \frac{2}{\Delta_{mn}(\textbf{k})} \frac{ \partial_i (E_{m\textbf{k}}+E_{n\textbf{k}})}{\Delta_{mn}(\textbf{k})-\hbar(\omega+i0^+)} \xi^j_{nm}\xi^l_{mn} 
\end{split}
\end{equation}
where the prime on the band index $s$ is to indicate that the sum is restricted such that $s\neq m, n$. Here we have simplified the contributions identified earlier \cite{PerryOptical}, as detailed in Appendix C, so that only off-diagonal components of the non-Abelian Berry connection matrix appear. Thus, with a knowledge of the band structure and cell periodic Bloch functions, the optical conductivity tensors can be computed with the sum over unoccupied bands truncated; for the off-diagonal non-Abelian Berry connection matrix elements can be replaced with the velocity matrix elements divided by the energy difference, avoiding complications at degenerate k-points in evaluating the non-Abelian Berry connection.

The tensor $\sigma^{il}(\omega)$ is symmetric in the presence of time reversal symmetry; more generally it is not. In the presence of time reversal symmetry the contributions to $\sigma^{ilj}(\omega)$ from $\alpha^{ij}_G$ vanish; more generally they are present. The additional frequency dependent contribution to $\sigma^{ilj}(\omega)$ is in general non-vanishing even with time-reversal symmetry, but does vanish in the presence of inversion symmetry.  

\end{widetext}

\section{Conclusions}\label{Conclusions}

In this paper we have extended a treatment of the optical response of crystals based on microscopic polarization and magnetization fields \cite{Perry_Sipe} to include electron spin. The Hamiltonian employed includes 
an interaction term between the spin and an external magnetic field, and spin-orbit coupling. Some of the real space expressions for quantities such as the ground state atomic and itinerant magnetization differ
from what was found previously \cite{PerryMagneto}, due to a
modified velocity operator and a different unperturbed Hamiltonian, but when converted to an integral over the Brillouin zone, they are formally identical to the previous work \cite{PerryMagneto}. The explicit expression for the spin magnetization is also identified.

In calculating the linear response we can generalize the ``orbital magnetoelectric polarizability" (OMP) tensors and construct magnetoelectric polarizability (MP) tensors that include spin contributions, both explicitly due to a Zeeman term in the Hamiltonian and implicitly in the spinor Bloch states used to calculate matrix elements. For the first time we give expressions for the MP tensors that include spin effects and are valid at finite frequencies; they reduce to a single tensor at zero frequency, even including spin effects. 

Other terms that contribute to the total current density induced by the electric field at finite frequency are also identified, leading to the third rank tensor, $\sigma^{ilj}(\omega)$, characterizing the response of the total current density to the spatial variation in the electric field. While individual contributions to the tensor are gauge-dependent, the full tensor itself is gauge invariant. This is in contrast to existing work on optical activity that begins with the total current response and then attempts to decompose the response into quadrupolar and magnetoelectric-like contributions, our decomposition is made apriori and it can be seen how the various contributions can arise and combine. This is the first time that an expression for this tensor, which characterizes optical activity, has been given that includes the effects of spin, holds at finite frequency, and is valid even if the crystal initially does not satisfy time-reversal symmetry. 

Finally, we have shown how to write the expression for the tensor $\sigma^{ilj}(\omega)$ in a form involving only off-diagonal components of the Berry connection. This makes the expression immediately suitable for numerical calculation.

\section{Acknowledgements}
We thank Jason Kattan and Perry Mahon for helpful comments and discussions.  This work was supported by the Natural Sciences and Engineering Research Council of Canada (NSERC).

\appendix
\begin{widetext}
\section{VELOCITY MATRIX ELEMENTS}\label{AppendixA}

To determine the form of the velocity matrix elements as functions of \textbf{k} we begin by manipulating the left side of equation (\ref{RealSpaceVel}),

\begin{equation}
\label{A1}
\begin{split}
    & \int \psi^\dag_{n'\textbf{k}'}(\textbf{x})\Big[\frac{\hbar}{im} \frac{\partial}{\partial x^a} - \frac{e}{mc} A^a_{stat}(\textbf{x}) + \frac{\hbar}{4m^2c^2} \epsilon^{abc} \sigma_b \frac{\partial \text{V}(\textbf{x})}{\partial x^c}\Big] \psi_{n\textbf{k}}(\textbf{x}) d\textbf{x}
    \\
    \\
    &= \frac{1}{(2\pi)^3} \int e^{i(\textbf{k}-\textbf{k}')\cdot\textbf{x}} K^a_{n'\textbf{k}'
    ;n\textbf{k}}(\textbf{x}) d\textbf{x},
\end{split}
\end{equation}
where
\begin{equation}
    K^a_{n'\textbf{k}';n\textbf{k}}(\textbf{x}) = u^\dag_{n'\textbf{k}'}(\textbf{x}) \Big[ \frac{\hbar}{im} \frac{\partial}{\partial x^a} + \frac{\hbar}{m} k^a - \frac{e}{mc}A^a_{stat}(\textbf{x})+ \frac{\hbar}{4m^2c^2} \epsilon^{abc} \sigma_b \frac{\partial \text{V}(\textbf{x})}{\partial x^c}\Big] u_{n\textbf{k}}(\textbf{x}).
\end{equation}
Then converting the integral over all space in equation (\ref{A1})  to a sum of integrals over unit cells leads to
\begin{equation}
\begin{split}
    &= \sum_{\textbf{R}} \frac{1}{(2\pi)^3} \int_{\Omega_{uc}} e^{i(\textbf{k}-\textbf{k}')\cdot\textbf{R}} e^{i(\textbf{k}-\textbf{k}')\cdot\textbf{y}} K^a_{n'\textbf{k}';n\textbf{k}}(\textbf{y}+\textbf{R}) d\textbf{y}
    \\
    &= \delta(\textbf{k}-\textbf{k}') \frac{1}{\Omega_{uc}} \int_{\Omega_{uc}} K^a_{n'\textbf{k};n\textbf{k}}(\textbf{y}) d\textbf{y}
    \\
    &= \delta(\textbf{k}-\textbf{k}') v^a_{n' n}(\textbf{k}),
\end{split}
\end{equation}
where
\begin{equation}
    v^a_{n' n}(\textbf{k}) = \frac{1}{\Omega_{uc}} \frac{1}{m} \int_{\Omega_{uc}} u^\dag_{n'\textbf{k}}(\textbf{x}) \left[
    \frac{\hbar}{i} \frac{\partial}{\partial x^a} + \hbar k^a - \frac{e}{c} A^a_{stat}(\textbf{x})+ \frac{\hbar}{4mc^2} \epsilon^{abc} \sigma_b \frac{\partial \text{V}(\textbf{x})}{\partial x^c}\right] u_{n\textbf{k}}(\textbf{x}) d\textbf{x}.
\end{equation}
To simplify the above integral we look at the effective Hamiltonian operator that acts on the cell-periodic part of the Bloch function

\begin{equation}
\begin{split}
    \mathcal{H}^0_\textbf{k}(\textbf{x}) = \frac{1}{2m}\left(
    \frac{\hbar}{i} \frac{\partial}{\partial x^b} - \frac{e}{c} A^b_{stat}(\textbf{x}) + \hbar k^b
    \right)^2 + V(\textbf{x}) + \frac{\hbar}{4m^2c^2} \epsilon^{abc} \sigma^b  \nabla_c \text{V}(\textbf{x}) &\left( \frac{\hbar}{i}\frac{\partial}{\partial x^a} - \frac{e}{c} A^a_{stat}(\textbf{x}) + \hbar k^a \right)
    \\
    &- \frac{e\hbar}{2mc} \sigma^a B^a_\text{static}(\textbf{x},t),
\end{split}
\end{equation}
and we take the derivative of the effective Hamiltonian with respect to \textbf{k},

\begin{equation}
   \frac{\partial}{\partial k^a} \mathcal{H}_\textbf{k}^0(\textbf{x}) = \frac{\hbar}{m} \left( \frac{\hbar}{i} \frac{\partial}{\partial x^a} - \frac{e}{c}A^a_{stat}(\textbf{x}) + \hbar k^a  + \frac{\hbar}{4mc^2} \epsilon^{abc} \sigma^b \nabla_c \text{V}(\textbf{x}) \right),
\end{equation}
which allows us to identify the velocity matrix element as

\begin{equation}
    v^a_{n'n}(\textbf{k}) = \frac{1}{\Omega_{uc}} \frac{1}{\hbar} \int_{\Omega_{uc}} u^\dag_{n'\textbf{k}}(\textbf{x}) \Big[ \frac{\partial }{\partial k^a} \mathcal{H}_\textbf{k}^0(\textbf{x}) \Big] u_{n\textbf{k}}(\textbf{x}) d\textbf{x}.
\end{equation}
The integral over the unit cell can be simplified 

\begin{equation}
\begin{split}
    \int_{\Omega_{uc}} u^\dag_{n'\textbf{k}}(\textbf{x}) \left[ \frac{\partial}{\partial k^a} \mathcal{H}_\textbf{k}^0(\textbf{x}) \right] u_{n\textbf{k}} d\textbf{x}
    \\
    = \frac{\partial}{\partial k^a} \int_{\Omega_{uc}} u^\dag_{n'\textbf{k}}(\textbf{x}) \mathcal{H}_\textbf{k}^0(\textbf{x}) u_{n\textbf{k}}(\textbf{x}) d\textbf{x} 
    - \int_{\Omega_{uc}} u^\dag_{n'\textbf{k}}(\textbf{x}) \mathcal{H}_\textbf{k}^0(\textbf{x}) \frac{\partial u_{n\textbf{k}}(\textbf{x})}{\partial k^a}  d\textbf{x}
    \\
    - \int_{\Omega_{uc}} \frac{\partial u^\dag_{n'\textbf{k}}(\textbf{x})}{\partial k^a} \mathcal{H}_\textbf{k}^0(\textbf{x}) u_{n\textbf{k}}(\textbf{x}) d\textbf{x}.
\end{split}
\end{equation}
Since

\begin{equation}
\begin{split}
    \mathcal{H}_\textbf{k}^0(\textbf{x}) u_{n\textbf{k}}(\textbf{x}) = E_{n\textbf{k}}u_{n\textbf{k}}(\textbf{x}),
\end{split}
\end{equation}
we get:

\begin{equation}
\begin{split}
    \delta_{n' n} \Omega_{uc} \frac{\partial}{\partial k^a} E_{n\textbf{k}} - E_{n'\textbf{k}} \int_{uc} u^\dag_{n'\textbf{k}}(\textbf{x}) \frac{\partial u_{n\textbf{k}}(\textbf{x})}{\partial k^a} d\textbf{x} - E_{n\textbf{k}} \int_{uc} \frac{\partial u^\dag_{n'\textbf{k}}(\textbf{x})}{\partial k^a} u_{n\textbf{k}}(\textbf{x}) d\textbf{x}
    \\
    = \Omega_{uc} \delta_{nn'} \frac{\partial E_{n\textbf{k}}}{\partial k^a} + \frac{1}{i}(E_{n\textbf{k}}-E_{n'\textbf{k}})\Omega_{uc}\xi^a_{n'n}(\textbf{k}).
\end{split}
\end{equation}
Putting this all together we find equation (\ref{kspaceVel}).
Thus we see that the expression for the \textbf{k} dependent velocity matrix elements of spinless electrons also holds when using spinors and including spin orbit coupling. 

\section{EXPRESSIONS FOR SITE QUANTITIES}\label{AppendixB}

The dynamics of the single particle density matrix $\eta_{\alpha\textbf{R};\beta\textbf{R}'}(t) \equiv \langle\hat{a}^\dag_{\beta\textbf{R}'}(t) \hat{a}_{\alpha\textbf{R}}(t) \rangle$ are governed by

\begin{equation}
\begin{split}
    i\hbar \frac{\partial \eta_{\alpha \textbf{R};\beta \textbf{R}'}(t)}{\partial t} = \sum_{\lambda \textbf{R}''} e^{i\Delta(\textbf{R},\textbf{R}'',\textbf{R}';t)}\big(\bar{H}_{\alpha\textbf{R};\lambda\textbf{R}''}(t)\eta_{\lambda\textbf{R}'';\beta\textbf{R}'}(t) 
    - \eta_{\alpha\textbf{R};\lambda\textbf{R}''}(t)\bar{H}_{\lambda\textbf{R}'';\beta\textbf{R}'}(t) 
    \big) - e\Omega^0_{\textbf{R}'}(\textbf{R};t)\eta_{\alpha\textbf{R};\beta\textbf{R}'}(t).
\end{split}
\end{equation}
The matrix $\bar{H}$ is Hermitian, and its matrix elements are

\begin{equation}
\label{HMatrix}
\begin{split}
    \bar{H}_{\alpha\textbf{R};\lambda\textbf{R}''}(t) =
    \frac{1}{2}  \int d\textbf{x}& \chi^*_{\alpha\textbf{R};\textsf{i}}(\textbf{x},t)e^{i\Delta(\textbf{R},\textbf{x},\textbf{R}'';t)}
    (\mathcal{K}_\textsf{ik}(\textbf{x},\textbf{R}'';t)\chi_{\lambda \textbf{R}'';\textsf{k}}(\textbf{x},t))
    \\
    + \frac{1}{2} \int d\textbf{x}& (\mathcal{K}_\textsf{ki}(\textbf{x},\textbf{R};t)\chi_{\alpha\textbf{R};\textsf{i}}(\textbf{x},t))^*e^{i\Delta(\textbf{R},\textbf{x},\textbf{R}'';t)}\chi_{\lambda \textbf{R}'';\textsf{k}}(\textbf{x},t)
    \\
    -\frac{i\hbar}{2}\int d\textbf{x}& e^{i\Delta(\textbf{R},\textbf{x},\textbf{R}'';t)} \Bigg( \chi^*_{\alpha \textbf{R};\textsf{i}}(\textbf{x},t) \frac{\partial \chi_{\lambda \textbf{R}'';\textsf{i}}(\textbf{x},t)}{\partial t} - 
    \frac{\partial \chi^*_{\alpha \textbf{R};\textsf{i}}(\textbf{x},t)}{\partial t}\chi_{\lambda \textbf{R}'';\textsf{i}}(\textbf{x},t)
    \\
    &-\frac{ie}{\hbar}\big(\Omega^0_{\textbf{R}''}(\textbf{x},t)+\Omega^0_\textbf{R}(\textbf{x},t)\big)\chi^*_{\alpha\textbf{R};\textsf{i}}(\textbf{x},t) \chi_{\lambda \textbf{R}'';\textsf{i}}(\textbf{x},t)
    \Bigg).
\end{split}
\end{equation}
\end{widetext}
Since all the terms appearing in the dynamics of $\eta_{\alpha\textbf{R}';\beta\textbf{R}''}$ are gauge-invariant in the electromagnetic sense, if its initial conditions are set before any fields are applied then it will remain so at all later times. The matrix elements $\bar{H}_{\alpha\textbf{R};\lambda\textbf{R}''}$ are effectively hopping matrix elements and on-site energies. While the matrix elements between different sites (\textbf{R}$\neq$\textbf{R}$''$) are in general non-zero, for \textbf{R} far from \textbf{R}$''$ they can be reasonably expected to be zero. Additionally, on the same site there are hopping matrix elements between different orbital types $\alpha$.

Isolating the contribution to the total charge density due to the electrons,
\begin{equation}
    \rho^e(\textbf{x},t) = \rho(\textbf{x},t) - \rho^\text{ion}(\textbf{x}),
\end{equation}
the electronic charge density $\rho^{e}(\textbf{x},t)$ can then be partitioned into contributions associated with the lattice sites \textbf{R}, 
\begin{equation}
    \rho^e(\textbf{x},t) = \sum_\textbf{R} \rho^e_\textbf{R}(\textbf{x},t) .
\end{equation}

The site electronic charge density and current densities are written in terms of the site Green function

\begin{equation}
\begin{split}
    \rho^e_\textbf{R}(\textbf{x},t) &= -ie\text{Tr}([G^\textbf{R}(\textbf{x},\textbf{y};t)]_{\textbf{y} \rightarrow \textbf{x}}) ,
    \\
    \textbf{j}_\textbf{R}(\textbf{x},t) &= \text{Tr}([\mathcal{J}^{\textbf{R}}(\textbf{x},\textbf{y};t)G^\textbf{R}(\textbf{x},\textbf{y};t)]_{\textbf{y}\rightarrow \textbf{x}}).
\end{split}
\end{equation}
We can re-express the microscopic charge density in terms of the single particle density matrix and a site quantity matrix

\begin{equation}
\label{electronicdensity}
\begin{split}
    \rho^e_{\textbf{R}}(\textbf{x},t) = \sum_{\alpha,\beta,\textbf{R}',\textbf{R}''}\rho_{\beta\textbf{R}';\alpha\textbf{R}''}(\textbf{x},\textbf{R};t)\eta_{\alpha\textbf{R}'';\beta\textbf{R}'},
\end{split}
\end{equation}
where
\begin{equation}
\begin{split}
    \rho_{\beta\textbf{R}';\alpha\textbf{R}''}(\textbf{x},\textbf{R};t) = \frac{e}{2} \left(\delta_{\textbf{R}\textbf{R}'} + \delta_{\textbf{R}\textbf{R}''}\right)
    \\
    \times  e^{i\Delta(\textbf{R}',\textbf{x},\textbf{R}'';t)}\chi^*_{\beta\textbf{R}';\textsf{i}}(\textbf{x},t)\chi_{\alpha\textbf{R}'';\textsf{i}}(\textbf{x},t).
\end{split}
\end{equation}

The total charge density associated with a site \textbf{R} is the sum of the electronic and nuclei contributions associated with said site ((\ref{electronicdensity}) and (\ref{nucleidensity}))

\begin{equation}
    \rho_\textbf{R}(\textbf{x},t) = \rho^e_\textbf{R}(\textbf{x},t) + \rho_\textbf{R}^\text{ion}(\textbf{x}).
\end{equation}

The charge current density is composed of two separate parts, one we wish to separately identify with the ``magnetization current'' that arises from the electron spin, as well as the more standard velocity charge current. 

\begin{equation}
\label{Jp}
\begin{split}
    \textbf{J}^{\mathfrak{p}}_\textsf{ij}(\textbf{x},\mathfrak{p}(\textbf{x},\textbf{y};t)) = \frac{e}{2m}\left(\boldsymbol{\mathfrak{p}}(\textbf{x})- \frac{e}{c}\Omega_\textbf{y}(\textbf{x},t)\right)\delta_\textsf{ij} 
    \\
    + \epsilon_{abc}\hat{e}^c \frac{e\hbar}{8m^2c^2}\sigma_\textsf{ij}^a\frac{ \partial  \text{V}(\textbf{x})}{\partial x_b} ,
\end{split}
\end{equation}

and

\begin{equation}
\label{Jm}
    \textbf{J}_\textsf{ij}^\textbf{M}(\textbf{x},\boldsymbol{\mathfrak{p}}(\textbf{x},\textbf{y};t)) = \epsilon_{abc} \hat{e}^c \frac{e\hbar}{2m} \sigma_\textsf{ij}^b \frac{\partial}{\partial x_a},
\end{equation}

and their sum

\begin{equation}
\label{JR}
    \textbf{J}^\textbf{R}_\textsf{ij}(\textbf{x},\mathfrak{p}(\textbf{x},\textbf{y};t)) = \textbf{J}_\textsf{ij}^{\mathfrak{p}}(\textbf{x},\mathfrak{p}(\textbf{x},\textbf{y};t) + \textbf{J}^\textbf{M}_\textsf{ij}(\textbf{x},\mathfrak{p}(\textbf{x},\textbf{y};t)).
\end{equation}

Using (\ref{Jp}), (\ref{Jm}), and (\ref{JR}) we can identify the microscopic charge current operator $\mathcal{J}^\textbf{R}$ as

\begin{equation}
    \mathcal{J}_\textsf{ji}^\textbf{R}(\textbf{x},\textbf{y};t) = -ie [\textbf{J}_\textsf{ji}^{\textbf{R}}(\textbf{x},\boldsymbol{\mathfrak{p}}(\textbf{x},\textbf{R};t))+\textbf{J}_\textsf{ji}^{\textbf{R}}(\textbf{y},\boldsymbol{\mathfrak{p}}^*(\textbf{y},\textbf{R};t))].
\end{equation}
This natural splitting of the current density allows one to identify two contributions to the expectation value of the site current

\begin{equation}
    \textbf{j}_\textbf{R}(\textbf{x},t) = \textbf{j}^{\mathfrak{p}}_\textbf{R}(\textbf{x},t) +  \textbf{j}^{\textbf{M}}_\textbf{R}(\textbf{x},t),
\end{equation}
where the microscopic velocity charge current is defined as

\begin{equation}
    \textbf{j}^{\mathfrak{p}}_\textbf{R}(\textbf{x},t) = \sum_{\alpha,\beta, \textbf{R}',\textbf{R}''} \textbf{j}^{\mathfrak{p}}_{\beta \textbf{R}';\alpha \textbf{R}''}(\textbf{x},\textbf{R};t) \eta_{\alpha \textbf{R}'';\beta \textbf{R}'}(t),
\end{equation}
with
\vspace{10pt}
\begin{widetext}

\begin{equation}
\begin{split}
    \textbf{j}^{\mathfrak{p}}_{\beta \textbf{R}';\alpha \textbf{R}''}(\textbf{x},\textbf{R};t) =& \Bigg(\frac{1}{2}\delta_{\textbf{R}\textbf{R}''}e^{i\Delta(\textbf{R}',\textbf{x},\textbf{R}'';t)}\chi^*_{\beta \textbf{R}'; \textsf{j}} (\textbf{x},t) \left(\textbf{J}^\mathfrak{p}_\textsf{ji}(\textbf{x},\boldsymbol{\mathfrak{p}}(\textbf{x},\textbf{R};t))\chi_{\alpha\textbf{R}'';\textsf{i}}(\textbf{x},t) \right)
    \\
    &+\frac{1}{2}\delta_{\textbf{R}\textbf{R}''}\left(\textbf{J}^\mathfrak{p}_\textsf{ji}(\textbf{x},\boldsymbol{\mathfrak{p}}^*(\textbf{x},\textbf{R};t))e^{i\Delta(\textbf{R}',\textbf{x},\textbf{R}'';t)}\chi^*_{\beta \textbf{R}' \textsf{j}}(\textbf{x},t)\right) \chi_{\alpha\textbf{R}''\textsf{i}}(\textbf{x},t) 
    \\
    &+\frac{1}{2}\delta_{\textbf{R}\textbf{R}'}\chi^*_{\beta \textbf{R}' \textsf{j}}(\textbf{x},t) \left(\textbf{J}^\mathfrak{p}_\textsf{ji}(\textbf{x},\boldsymbol{\mathfrak{p}}(\textbf{x},\textbf{R};t)) e^{i\Delta(\textbf{R}',\textbf{x},\textbf{R}'';t)}\chi_{\alpha\textbf{R}''\textsf{i}}(\textbf{x},t) \right) 
    \\
    &+\frac{1}{2}\delta_{\textbf{R}\textbf{R}'}\left(\textbf{J}^\mathfrak{p}_\textsf{ji}(\textbf{x},\boldsymbol{\mathfrak{p}}^*(\textbf{x},\textbf{R};t))\chi^*_{\beta \textbf{R}' \textsf{j}}(\textbf{x},t)\right)e^{i\Delta(\textbf{R}',\textbf{x},\textbf{R}'';t)} \chi_{\alpha\textbf{R}''\textsf{i}}(\textbf{x},t) 
    \Bigg).
\end{split}
\end{equation}
As well the magnetization current is given by

\begin{equation}
    \textbf{j}^{\textbf{M}}_{\textbf{R}}(\textbf{x},t) = \sum_{\alpha,\beta,\textbf{R}',\textbf{R}''} \textbf{j}^{\textbf{M}}_{\beta\textbf{R}';\alpha\textbf{R}''}(\textbf{x},\textbf{R};t)\eta_{\alpha\textbf{R}'';\beta\textbf{R}'}(t),
\end{equation}
with

\begin{equation}
\begin{split}
    \textbf{j}^{\textbf{M}}_{\beta\textbf{R}';\alpha\textbf{R}''}(\textbf{x},\textbf{R};t) =
    \left(\delta_{\textbf{R}\textbf{R}'} + \delta_{\textbf{R}\textbf{R}''}\right) \frac{e\hbar}{4m} \nabla \times 
    \left(   e^{i\Delta(\textbf{R}',\textbf{x},\textbf{R}'')} \chi^*_{\beta \textbf{R}';\textsf{j}}(\textbf{x},t)  \boldsymbol\sigma_\textsf{ji} \chi_{\alpha \textbf{R}'';\textsf{i}}(\textbf{x},t) \right).
\end{split}
\end{equation}

\end{widetext}

The expressions for the site specific current and charge density do not in general satisfy continuity,

\begin{equation}
    K_\textbf{R}(\textbf{x},t) \equiv \nabla \cdot \textbf{j}_\textbf{R}(\textbf{x},t) + \frac{\partial \rho_\textbf{R}(\textbf{x},t)}{\partial t} \neq 0.
\end{equation}
This is because charges are free to travel from a region associated with site \textbf{R} to other regions. The total charge associated with a region \textbf{R} is in general a time-dependent quantity. However, since total charge is conserved the sum over all $K_\textbf{R}$ is zero. Note that the divergence of the magnetization current is always zero since the magnetization current is the curl of a vector. 

The total site charges are 

\begin{equation}
\label{esitecharg}
    Q_\textbf{R}(t) \equiv \int \rho_\textbf{R}(\textbf{x},t) d\textbf{x} = e\sum_\alpha \eta_{\alpha \textbf{R};\alpha \textbf{R}}(t) + \sum_N q_N,
\end{equation}
which follows from how we defined $\rho_\textbf{R}(\textbf{x},t)$ and our othonormality condition for the Wannier functions. Link currents are introduced since the total charge associated with site \textbf{R} is not necessarily constant in time

\begin{equation}
\label{linkcurrents}
\begin{split}
    \frac{dQ_\textbf{R}(t)}{dt} = \sum_{\textbf{R}'} I(\textbf{R},\textbf{R}';t), 
\end{split}
\end{equation}
with
\begin{equation}
\begin{split}
    I(\textbf{R},\textbf{R}';t)= \frac{e}{i\hbar} \sum_{\alpha, \lambda}\Big( 
    \bar{H}_{\alpha \textbf{R};\lambda \textbf{R}'}(t) \eta_{\lambda\textbf{R}';\alpha\textbf{R}}(t)
    \\
    - \eta_{\alpha \textbf{R};\lambda \textbf{R}'}(t) \bar{H}_{\lambda \textbf{R}';\alpha \textbf{R}}(t)
    \Big).
\end{split}
\end{equation}

With the total site charges and links currents identified in (\ref{esitecharg}) and (\ref{linkcurrents}), we can now define the microscopic ``free" charge and current densities

\begin{equation}
\label{FreeCharge}
    \rho_F(\textbf{x},t) \equiv \sum_\textbf{R}\delta(\textbf{x}-\textbf{R}) Q_\textbf{R}(t),
\end{equation}
and
\begin{equation}
\label{FreeCurrent}
    \textbf{j}_F(\textbf{x},t) \equiv \frac{1}{2} \sum_{\textbf{R},\textbf{R}'} \textbf{s}(\textbf{x};\textbf{R},\textbf{R}')I(\textbf{R},\textbf{R}';t).
\end{equation}
The free charge density is simply the sum of the charges associated with each lattice site placed at that lattice site. The second introduces a microscopic current density by distributing the net link current along a path from site \textbf{R}$'$ to \textbf{R}. The factor of 1/2 is so one can sum over all \textbf{R} and \textbf{R}$'$, compensating for double counting. These expressions satisfy continuity

\begin{equation}
    \nabla \cdot \textbf{j}_F(\textbf{x},t) + \frac{\partial \rho_F(\textbf{x},t)}{\partial t} = 0.
\end{equation}

As can be seen in equation (\ref{SiteMagnetizations}), the magnetization has been split into three contributions. These contributions are termed the ``atomic", ``itinerant", and ``spin" contribution. The atomic magnetization is defined as

\begin{equation}
    \bar{m}^j_\textbf{R}(\textbf{x},t) \equiv \frac{1}{c} \int \alpha^{jk}(\textbf{x};\textbf{y},\textbf{R})j^{\mathfrak{p},k}_\textbf{R}(\textbf{y},t) d\textbf{y}.
\end{equation}
The itinerant magnetization is included due to the possibility of charges being allowed to move between sites. The derivation is fairly involved, and we refer the reader to earlier work \cite{Perry_Sipe}. The relevant equations are

\begin{equation}
    \tilde{m}_\textbf{R}^j(\textbf{x},t) \equiv \frac{1}{c} \int \alpha^{jk}(\textbf{x};\textbf{y},\textbf{R})\tilde{j}^k_\textbf{R}(\textbf{y},t) d\textbf{y},
\end{equation}
where
\begin{equation}
    \tilde{\textbf{j}}_\textbf{R}(\textbf{x},t) = \sum_{\alpha,\beta,\textbf{R}',\textbf{R}''} \tilde{\textbf{j}}_{\beta\textbf{R}';\alpha\textbf{R}''}(\textbf{x},\textbf{R};t) \eta_{\alpha\textbf{R}'';\beta\textbf{R}'}(t).
\end{equation}
With
\begin{equation}
    \tilde{\textbf{j}}_{\beta\textbf{R}';\alpha\textbf{R}''}(\textbf{x},\textbf{R};t) = \frac{1}{2}(\delta_{\textbf{R}\textbf{R}''} + \delta_{\textbf{R}\textbf{R}'}) \tilde{\textbf{j}}_{\beta\textbf{R}';\alpha\textbf{R}''}(\textbf{x},t),
\end{equation}
and
\begin{equation}
\begin{split}
    \tilde{\textbf{j}}_{\beta \textbf{R}''';\alpha \textbf{R}''}(\textbf{x},t) =& -\sum_\textbf{R} \int s(\textbf{x};\textbf{y},\textbf{R})\Gamma_\textbf{R}^{\alpha\textbf{R}'';\beta\textbf{R}'''}(\textbf{y},t) d\textbf{y}
    \\
    &- \frac{1}{2}\sum_{\textbf{R},\textbf{R}'} s(\textbf{x};\textbf{R},\textbf{R}') \zeta_{\textbf{R}\textbf{R}'}^{\alpha\textbf{R}'';\beta\textbf{R}'''}(t),
\end{split}
\end{equation}
where
\begin{equation}
\begin{split}
    \zeta_{\textbf{R}\textbf{R}'}^{\alpha\textbf{R}'';\beta\textbf{R}'''}(t) = \frac{e}{i\hbar}\Big(
    \delta_{\textbf{R}'''\textbf{R}}\delta_{\textbf{R}''\textbf{R}'}\bar{H}_{\beta\textbf{R};\alpha\textbf{R}'}(t) 
    \\
    - \delta_{\textbf{R}''\textbf{R}}\delta_{\textbf{R}'''\textbf{R}'}\bar{H}_{\beta\textbf{R}';\alpha\textbf{R}}(t)
    \Big),
\end{split}
\end{equation}
and 
\begin{equation}
    \begin{split}
        \Gamma_\textbf{R}^{\alpha\textbf{R}'';\beta\textbf{R}'}(\textbf{x},t) = \nabla \cdot \textbf{j}^{\mathfrak{p}}_{\beta \textbf{R}';\alpha \textbf{R}''}(\textbf{x},\textbf{R};t)
        \\
        + \frac{\partial \rho_{\beta\textbf{R}';\alpha\textbf{R}''}(\textbf{x},\textbf{R};t)}{\partial t} 
        \\
        + \frac{1}{i\hbar} \sum_{\mu,\nu,\textbf{R}_1,\textbf{R}_2}\rho_{\nu\textbf{R}_2;\mu\textbf{R}_1}(\textbf{x},\textbf{R};t)\mathfrak{F}^{\alpha\textbf{R}'';\beta\textbf{R}'}_{\mu\textbf{R}_1;\nu \textbf{R}_2}(t),
    \end{split}
\end{equation}
with
\begin{equation}
\begin{split}
    \mathfrak{F}^{\alpha\textbf{R}'';\beta\textbf{R}'}_{\mu\textbf{R}_1;\nu\textbf{R}_2}(t) = \delta_{\beta\nu}\delta_{\textbf{R}_2\textbf{R}'}e^{i\Delta(\textbf{R}_1,\textbf{R}'',\textbf{R}_2;t)}\bar{H}_{\mu\textbf{R}_1;\alpha\textbf{R}''}(t) \\
    -\delta_{\alpha\mu} \delta_{\textbf{R}''\textbf{R}_1}e^{i\Delta(\textbf{R}_1,\textbf{R}',\textbf{R}_2;t)}\bar{H}_{\beta\textbf{R}';\nu\textbf{R}_2}(t) \\
    - e\delta_{\beta\nu}\delta_{\alpha\mu}\delta_{\textbf{R}_2\textbf{R}'}\delta_{\textbf{R}_1\textbf{R}''}\Omega^0_{\textbf{R}_2}(\textbf{R}_1;t).
\end{split}
\end{equation}

The spin contribution to the magnetization is given by

\begin{equation}
\begin{split}
    \breve{\textbf{m}}(\textbf{x},t) = \frac{e\hbar}{2mc}
    \langle \psi^\dag(\textbf{x},t) \boldsymbol\sigma \psi(\textbf{x},t) \rangle
    \\
    = \frac{e\hbar}{4mc} \sum_{\alpha,\beta,\textbf{R},\textbf{R}',\textbf{R}''} \left( \delta_{\textbf{R}\textbf{R}'} + \delta_{\textbf{R}\textbf{R}''}\right) \eta_{\alpha\textbf{R}'';\beta\textbf{R}'}(t)
    \\
    \times e^{i\Delta(\textbf{R}',\textbf{x},\textbf{R}'')} \chi^*_{\beta\textbf{R}';\textsf{i}}(\textbf{x},t)\boldsymbol\sigma_\textsf{ij}\chi_{\alpha \textbf{R}'';\textsf{j}}(\textbf{x},t),
\end{split}
\end{equation}
which can be broken up into site magnetizations in an obvious way,
\begin{equation}
    \breve{\textbf{m}}(\textbf{x},t) = \sum_\textbf{R} \breve{\textbf{m}}_\textbf{R}(\textbf{x},t).
\end{equation}
The expression for $\breve{\textbf{m}}_\textbf{R}(\textbf{x},t)$ has already been given in Section \ref{MicroFormalism}, equation (\ref{SpinMag}).

The total microscopic polarization field is just the sum of all the site polarizations, and likewise the total microscopic magnetization field is just the sum of all the site magnetizations. This follows what was done earlier \cite{Perry_Sipe} when the formalism was outlined excluding electron spin. The key differences are that now the Wannier functions are spinors, the velocity operator is modified by spin-orbit coupling, and there is an extra contribution to the magnetization explicitly due to the spin.

\section{Effective Conductivity Tensor}\label{AppendixC}

The effective third-rank conductivity tensor $\sigma^{ilj}(\omega)$ obtained by our formalism is naturally decomposed into portions associated with the electric quadrupole and magnetic dipole response to an applied electric field, and the electric dipole response to a magnetic field and spatial derivatives of the electric field. These contributions all combine \cite{PerryOptical} to produce the total gauge-invariant response

\begin{equation}
\begin{split}
    \sigma^{ilj}(\omega) =& \omega \breve{\gamma}^{ijl}(\omega) - \omega \breve{\chi}_Q^{ijl}(\omega)
    \\
    &-ic\breve{\alpha}^{ia}_{\mathcal{P}}(\omega) \epsilon^{ajl} + ic \epsilon^{ijb}\breve{\alpha}^{lb}_{\mathcal{M}}(\omega).
\end{split}
\end{equation}
The gauge-invariant response tensors denoted with a breve accent are

\begin{widetext}

\begin{equation}
    \breve{\gamma}^{ijl}(\omega) \equiv \frac{e^2}{4} \sum_{mn} f_{nm} \int_{BZ} \frac{d\textbf{k}}{(2\pi)^3} \Big[ \mathcal{F}^{jl}_{mn}(\textbf{k},\omega) + \mathcal{F}^{lj}_{mn}(\textbf{k},\omega) \Big] \xi^i_{nm},
\end{equation}
\begin{equation}
    \breve{\chi}_Q^{ijl}(\omega) \equiv \frac{e^2}{4} \sum_{mns} f_{nm} \int_{BZ} \frac{d\textbf{k}}{(2\pi)^3} \frac{ \xi^l_{mn}(\xi^i_{ns}\xi^j_{sm} + \xi^j_{ns}\xi^i_{sm} ) }{\Delta_{mn}(\textbf{k})-\hbar(\omega+i0^+)},
\end{equation}
\begin{equation}
    \breve{\alpha}^{ia}_\mathcal{P}(\omega) \equiv \alpha^{ia}_{G} + \alpha^{ia}_{S;\mathcal{P}}(\omega) +  \frac{\omega e^2}{4c} \epsilon^{abc} \sum_{mn} f_{nm} \int_{BZ} \frac{d\textbf{k}}{(2\pi)^3} \frac{ \grave{B}^{bc}_{mn}(\textbf{k},\omega) \xi^i_{nm} }{\Delta_{mn}(\textbf{k})-\hbar(\omega+i0^+)},
\end{equation}
where
\begin{equation}
    \grave{B}^{bc}_{mn}(\textbf{k},\omega) \equiv i\sum_s \Bigg[ \frac{\Delta_{sn}(\textbf{k})}{\Delta_{mn}(\textbf{k})} \xi^b_{ms} \xi^c_{sn} + \frac{ \Delta_{sm}(\textbf{k})}{\Delta_{mn}(\textbf{k})} \xi^b_{ms}\xi^c_{sn} \Bigg] -\Bigg[ 3 + \frac{\hbar\omega}{\Delta_{mn}(\textbf{k})-\hbar(\omega+i0^+)} \Bigg] \frac{\partial_b (E_{m\textbf{k}}+E_{n\textbf{k}})}{\Delta_{mn}(\textbf{k})} \xi^c_{mn},
\end{equation}
and lastly
\begin{equation}
\begin{split}
    \breve{\alpha}_\mathcal{M}^{lb}(\omega) \equiv  &\alpha^{lb}_{G} + \alpha^{lb}_{S;\mathcal{M}}(\omega)
    + \frac{\omega e^2}{4c} \epsilon^{bac} \sum_{mn} f_{nm} \int_{BZ} \frac{d\textbf{k}}{(2\pi)^3} \frac{1}{\Delta_{mn}(\textbf{k})-\hbar(\omega+i0^+)}
    \\
    &\times \Bigg[ 2 \frac{ \partial_c(E_{m\textbf{k}}+E_{n\textbf{k}})}{\Delta_{mn}(\textbf{k})} \xi^a_{nm} \xi^l_{mn} + i \sum_s \Bigg[ \frac{\Delta_{sm}(\textbf{k})}{\Delta_{mn}(\textbf{k})}  - \frac{\Delta_{ns}(\textbf{k})}{\Delta_{mn}(\textbf{k})} \Bigg] \xi^a_{ns}\xi^c_{sm} \xi^l_{mn} \Bigg].
\end{split}
\end{equation}

The Chern-Simons contribution to the magneto-polarizability plays no role in describing optical activity so is not present in $\breve{\alpha}^{ia}_{\mathcal{P}} $ and $\breve{\alpha}^{lb}_{\mathcal{M}}$. We can identify the portion of $\sigma^{ilj}(\omega)$ that remains when treating the zero frequency response as

\begin{equation}
    \sigma^{ilj}_{DC} = -ic\alpha_{G}^{ia}\epsilon^{ajl} + ic \epsilon^{ijb} \alpha^{lb}_{G} -ic\alpha^{ia}_{\mathcal{S};\mathcal{P}}(0) \epsilon^{ajl} + ic\epsilon^{ijb} \alpha^{lb}_{\mathcal{S};\mathcal{M}}(0),
\end{equation}
since at zero frequency all other contributions vanish; $\sigma^{ilj}_{DC}$ vanishes if the system does not break both time-reversal and inversion symmetry. On the other hand, the frequency dependent response only requires inversion symmetry to be broken.

We can write $\sigma^{ilj}(\omega)$ as

\begin{equation}
\label{sigmaPreStep}
\begin{split}
    \sigma^{ilj}(\omega) = -ic\alpha_{G}^{ia}\epsilon^{ajl} + ic \epsilon^{ijb}& \alpha^{lb}_{G} -ic\alpha^{ia}_{\mathcal{S};\mathcal{P}}(\omega) \epsilon^{ajl} + ic\epsilon^{ijb} \alpha^{lb}_{\mathcal{S};\mathcal{M}}(\omega)
    \\
    +\frac{ \omega e^2}{4} \sum_{mns} f_{nm} &\int \frac{d\textbf{k}}{(2\pi)^3} \Bigg[ \frac{ \xi^j_{ms}\xi^l_{sn} \xi^i_{nm} + \xi^l_{ms}\xi^j_{sn} \xi^i_{nm} }{ \Delta_{mn}(\textbf{k}) -\hbar(\omega+i0^+) } 
    + i\frac{ \partial_j (E_{m\textbf{k}}+E_{n\textbf{k}}) \xi^l_{mn} + \partial_l (E_{m\textbf{k}}+E_{n\textbf{k}})\xi^j_{mn} }{(\Delta_{mn}(\textbf{k})-\hbar(\omega+i0^+))^2} \xi^i_{nm} \Bigg]
    \\
    -\frac{\omega e^2}{4} \sum_{mns} f_{nm} &\int_{BZ} \frac{d\textbf{k}}{(2\pi)^3} \frac{ \xi^l_{mn}(\xi^i_{ns}\xi^j_{sm}+\xi^j_{ns}\xi^i_{sm})}{\Delta_{mn}(\textbf{k}) -\hbar(\omega+i0^+)}
    \\
    -i\epsilon^{ajl}\epsilon^{abc} \frac{\omega e^2}{4} &\sum_{mns} f_{nm} \int_{BZ} \frac{d\textbf{k}}{(2\pi)^3} \frac{\xi^i_{nm}}{\Delta_{mn}(\textbf{k})-\hbar(\omega+i0^+)} \Bigg[
    i\frac{\Delta_{sn}(\textbf{k}) + \Delta_{sm}(\textbf{k}) }{\Delta_{mn}(\textbf{k})} \xi^b_{ms}\xi^c_{sn} 
    \\
    &-\Bigg(3+ \frac{\hbar\omega}{\Delta_{mn}(\textbf{k})-\hbar(\omega+i0^+)}\Bigg) \frac{ \partial_b (E_{m\textbf{k}}+E_{n\textbf{k}}) }{\Delta_{mn}(\textbf{k})} \xi^c_{mn}
    \Bigg]
    \\
    +i \epsilon^{ijb} \epsilon^{bac} \frac{\omega e^2}{4} &\sum_{mns} f_{nm} \int_{BZ} \frac{d\textbf{k}}{(2\pi)^3} \frac{1}{\Delta_{mn}(\textbf{k}) -\hbar(\omega+i0^+)} \Bigg[ i\frac{ \Delta_{sm}(\textbf{k}) + \Delta_{sn}(\textbf{k}) }{\Delta_{mn}(\textbf{k})}
    \xi^a_{ns}\xi^c_{sm}\xi^l_{mn}
    \\
    &+ 2 \frac{ \partial_c (E_{m\textbf{k}}+E_{n\textbf{k}})}{\Delta_{mn}(\textbf{k})} \xi^a_{nm}\xi^l_{mn}
    \Bigg]
\end{split}
\end{equation}
Using the identity for the product of two levi-civita tensors with a shared index $\epsilon^{ajl}\epsilon^{abc} = \delta_{jb}\delta_{lc} - \delta_{jc}\delta_{lb}$ and $\epsilon^{ijb}\epsilon^{bac} = \delta_{ia}\delta_{jc} - \delta_{ja}\delta_{ic}$ we can then rewrite equation (\ref{sigmaPreStep}) as equation (\ref{effectiveqlinearconductivity}). Focusing on the second and third lines of equation (\ref{effectiveqlinearconductivity}) we see that these are the only lines that seemingly contain diagonal Berry connection matrix elements 
\begin{equation}
\label{DiagonalElementsOptoCond}
\begin{split}
    &\frac{\omega e^2}{4} \sum_{mns} f_{nm} \int_{BZ} \frac{d\textbf{k}}{(2\pi)^3} \frac{1}{\Delta_{mn}(\textbf{k}) -\hbar(\omega+i0^+)} 
    \Bigg( \xi^i_{nm}\xi^j_{ms}\xi^l_{sn} - \xi^i_{ns}\xi^j_{sm}\xi^l_{mn}\Bigg)
    \Bigg[ 1 + \frac{\Delta_{sn}(\textbf{k}) - \Delta_{ms}(\textbf{k})}{\Delta_{mn}(\textbf{k})}\Bigg]
    \\
    &+\frac{\omega e^2}{4} \sum_{mns} f_{nm} \int_{BZ} \frac{d\textbf{k}}{(2\pi)^3} \frac{1}{\Delta_{mn}(\textbf{k}) -\hbar(\omega+i0^+)} 
    \Bigg( \xi^i_{nm}\xi^l_{ms}\xi^j_{sn}
    -\xi^i_{sm}\xi^l_{mn}\xi^j_{ns} \Bigg)
    \Bigg[ 1 - \frac{\Delta_{sn}(\textbf{k}) - \Delta_{ms}(\textbf{k})}{\Delta_{mn}(\textbf{k})}\Bigg].
\end{split}
\end{equation}
However, written in this way we make it apparent that these diagonal elements cannot contribute to optical activity so do not need to be included in the sum. In the first line of equation ($\ref{DiagonalElementsOptoCond}$) if s=n the term in square brackets vanishes, if s=m then the term in round brackets vanishes. Likewise, in the second line of equation ($\ref{DiagonalElementsOptoCond}$) if s=m the term in square brackets vanishes, and if s=n the term in round brackets vanishes. Thus the sum over s should exclude n and m, avoiding the need to calculate the diagonal elements of the Berry connection which cannot be expressed in terms of the velocity matrix elements. 

\end{widetext}

\section{Symmetry Constraints on MP Tensor}\label{AppendixD}
In this appendix we confirm that the zero frequency MP tensor vanishes unless both time reversal and inversion symmetry are broken, with the caveat that the gauge-dependent part of the MP tensor, entirely contained in the Chern-Simons contribution, does not necessarily vanish. The frequency dependent MP tensor only requires inversion symmetry breaking. For convenience we include the form of the Chern-Simons and cross-gap contribution. 
\begin{widetext}

The Chern-Simons contribution can be written as \cite{PerryMagneto}  
\begin{equation}
\begin{split}
    \alpha^{il}_{CS} =-\delta^{il} \frac{e^2}{2\hbar c} \epsilon^{abc} \int_{BZ} \frac{d\textbf{k}}{(2\pi)^3} \sum_{vv'} \Bigg[ \Bigg(  \xi^a_{vv'} \partial_b \xi^c_{v' v} -\frac{2i}{3} \sum_{v_1} \xi^a_{vv'} \xi^b_{v'v_1} \xi^c_{v_1 v} \Bigg) + (\partial_b \mathcal{W}^a_{vv'}) \mathcal{W}^c_{v'v} - \frac{2i}{3} \sum_{v_1} \mathcal{W}^a_{vv'} \mathcal{W}^b_{v'v_1} \mathcal{W}^c_{v_1 v} \Bigg].
\end{split}
\end{equation}
This is the usual Chern-Simons contribution to the MP tensor \cite{VanderbiltOMP, OMP1, OMP2, OMP3}. The ``cross gap" contribution can be written as \cite{PerryMagneto}

\begin{equation}
\label{alphaG}
\begin{split}
    \alpha^{il}_G = \frac{e^2}{\hbar c}\epsilon^{lab} \int_{BZ} \frac{d\textbf{k}}{(2\pi)^3} \Bigg\{  -&\sum_{cv} \frac{ \partial_b (E_{c\textbf{k}}+E_{v\textbf{k}})}{E_{v\textbf{k}}-E_{c\textbf{k}}} \text{Re}\Big[ (\partial_a v| c) (c | \partial_i v) \Big]
    -\sum_{vv'c} \frac{ E_{v\textbf{k}} - E_{v'\textbf{k}}}{E_{v\textbf{k}} - E_{c\textbf{k}}} \text{Re}\Big[ (\partial_b v | v' ) (\partial_a v'|c) (c|\partial_i v) \Big] 
    \\
    +&\sum_{vcc'} \frac{ E_{c\textbf{k}} - E_{c'\textbf{k}}}{E_{v\textbf{k}} - E_{c\textbf{k}}} \text{Re}\Big[ (\partial_b v| c')(c'|\partial_a c)(c |\partial_i v) \Big]
    \Bigg\},
\end{split}
\end{equation}

\end{widetext}
where $v$ labels an occupied state and $c$ labels an unoccupied state. When $n$, $m$, or $s$ are used the index runs over both occupied and unoccupied states, like in the spin contribution equation (\ref{alphaS_Tensor}). The Chern-Simons and cross gap contributions have not changed their form by adding the spin degree of freedom, the Zeeman term, and spin-orbit coupling. From the form of equation \ref{alphaG} it is clear that the ``degeneracy" and ``reflection" conditions outlined in Essin et al \cite{VanderbiltOMP} cause $\alpha_G$ to vanish.  

\subsection{Inversion symmetry - Bloch functions}

In the case of inversion symmetry one requires that the Hamiltonian satisfies $\mathcal{H}^{0}(\textbf{x}) = \mathcal{H}^{0}(-\textbf{x})$. This implies that the eigenstate $\psi_{n\textbf{k}}(\textbf{x})$ has the same energy as the state $\psi_{n\textbf{k}}(-\textbf{x})$. Furthermore, the new state $\psi_{n\textbf{k}}(-\textbf{x})$ can in fact be identified with a Bloch state with wavevector -\textbf{k} \cite{CallawaySolidState}. Thus, we may choose the cell-periodic Bloch spinors at opposite momenta to be identical, 

\begin{equation}
\label{SISBloch}
    u_{n}(\textbf{k}) = u_{n}(-\textbf{k}).
\end{equation}
The relationship in equation (\ref{SISBloch}) could be altered by a \textbf{k} dependent phase factor. This represents an alternative gauge choice, only affecting the diagonal elements of U(\textbf{k}), seen in equation (\ref{WannierKet}). 

Choosing Bloch functions that satisfy (\ref{SISBloch}) the non-Abelian Berry connection at opposite momenta is related by

\begin{equation}
\label{InversionBerryConnection}
    \xi^a_{nm}(-\textbf{k}) = - \xi^a_{nm}(\textbf{k}).
\end{equation}
As well the spin matrix elements at opposite momenta are the same

\begin{equation}
\label{InversionSpinElements}
    S^i_{nm}(-\textbf{k}) =  S_{nm}^i(\textbf{k}).
\end{equation}
Now the question becomes what consequences these constraints have on the MP tensor. We can only make a definitive statement on the gauge-independent part of the MP. Split the integrals in half and perform a change of variables of \textbf{k} $\rightarrow$ -\textbf{k}. The result is as one would expect zero, bar the gauge-dependent terms. Had we chosen the Bloch functions to be related by some \textbf{k}-dependent phase with a non-zero derivative the relationships (\ref{InversionBerryConnection}) and (\ref{InversionSpinElements}) would be more complicated but the result for the MP tensor would be the same. This is because the MP tensor is insensitive to purely diagonal gauge transformations. This vanishing of $\alpha^{il}_S$ and $\alpha^{il}_G$ for zero frequency holds also at finite frequency. The same arguments lead to $\alpha^{li}_\mathcal{M}(\omega)$ and $\alpha^{il}_\mathcal{P}(\omega)$ vanishing with inversion symmetry (Neglecting the gauge-dependence).  

\subsection{Time reversal - Bloch functions}

The time reversal operator for half-integer spinors can be written as \cite{ManoukianBook,GroupTheoryDresselhaus}

\begin{equation}
    \mathcal{T} \rightarrow i\sigma_y \mathcal{K},
\end{equation}
where $\sigma_y$ is the second of the Pauli matrices, $\mathcal{K}$ is the complex conjugation operator, and we use the arrow to denote the spinor representation of the operator. A general Bloch function can be labelled by its energy eigenvalue $n$ and crystal momentum \textbf{k}, as is shown in equation (\ref{Blochfunctions}). We can easily examine the effect of the time reversal operator on a general Bloch state 
\begin{equation}
    \mathcal{T} \psi_{n\textbf{k}}(\textbf{x}) = \frac{ e^{-i\textbf{k}\cdot\textbf{x}}}{(2\pi)^{3/2}} 
    \begin{bmatrix}
    u^*_{n,\textbf{k}\downarrow}(\textbf{x})
    \\
    -u^*_{n,\textbf{k}\uparrow}(\textbf{x})
    \end{bmatrix}.
\end{equation}
If the time reversal operator and the Hamiltonian commute then $\mathcal{T}\psi_{n\textbf{k}}(\textbf{x})$ is an eigenstate with the same energy as $\psi_{n\textbf{k}}(\textbf{x})$, but is orthogonal. Let us label this state by $n'$ instead of $n$, to indicate that it is distinct but has the same energy. The time-reversed state also is an eigenstate of the translation operator with eigenvalue -\textbf{k} instead of \textbf{k}. Thus we can write

\begin{equation}
    \mathcal{T}|\psi_{n\textbf{k}}\rangle =  e^{i\phi_n(\textbf{k})} |\psi_{n'-\textbf{k}} \rangle,
\end{equation}
where there is the possibility of an arbitrary \textbf{k} dependent phase relating the eigenstates. The spin expectation value of $\psi_{n\textbf{k}}$ is the negative of $\psi_{n'-\textbf{k}}$.

We also have the relationship between the energy of the time reversed states \cite{GroupTheoryDresselhaus}, 

\begin{equation}
    E_{n}(\textbf{k}) = E_{n'}(-\textbf{k}).
\end{equation}

When we sum over all states we will have time reversed partners appearing. We assume the ground state is time-reversal symmetric, so if a state is filled in the ground state its time-reversed partner is also filled. The relationships the non-Abelian Berry connection and spin matrix elements obey, complicated by the addition of primed indices due to Kramer's pairs, are:

\begin{equation}
\begin{split}
    \xi^a_{nm}(\textbf{k})&= \xi^a_{m'n'}(-\textbf{k})
    \\
    S^a_{nm}(\textbf{k}) &= -S^a_{m'n'}(-\textbf{k})
    \\
    \xi^a_{nm'}(\textbf{k}) &= -\xi^a_{mn'}(-\textbf{k})
    \\
    S^a_{nm'}(\textbf{k}) &= S^a_{mn'}(-\textbf{k})
\end{split}
\end{equation}

If one takes the integrals in the MP and splits them in half, performs a change of variables from \textbf{k} to -\textbf{k} in one of them, one will find that all that remains are the gauge-dependent terms.

This cancellation does not occur for the finite frequency case. This is because a change of indices $n\leftrightarrow m$ is required in the above procedure, the energy denominator $\Delta_{nm}(\textbf{k})$ will then merely pick up a negative sign upon swapping band indices. In the finite frequency case the denominator $\Delta_{nm}(\textbf{k})-\hbar(\omega+i0^+)$ appears, which does not have that simple relationship upon swapping band indices. However, the cancellation can still occur at finite frequency if one begins with an identical Hamiltonian for the spin-up and spin-down electrons, the `scalar particle' treatment, and uses $\mathcal{T} \rightarrow \mathcal{K}$. Then $\psi_{n\textbf{k}}(\textbf{x}) = \psi^*_{n-\textbf{k}}(\textbf{x})$, $E_{n}(\textbf{k}) = E_{n}(-\textbf{k})$, $\psi_{n\textbf{k}\uparrow}(\textbf{x}) = \begin{bmatrix} \psi_{n\textbf{k}}(\textbf{x}), & 0
\end{bmatrix}$, $\psi_{n\textbf{k}\downarrow}(\textbf{x}) = \begin{bmatrix} 0, & \psi_{n\textbf{k}}(\textbf{x})
\end{bmatrix}$, and $E_{n\uparrow}(\textbf{k}) = E_{n\downarrow}(\textbf{k}) = E_{n}(\textbf{k})$. The inclusion of spin-orbit coupling, while not breaking time-reversal symmetry, can be sufficient to lift this degeneracy and lead to a non-zero spin contribution to the MP tensor.

\bibliographystyle{apsrev4-2}
\bibliography{references.bib}

\begin{thebibliography}{51}%
\makeatletter
\providecommand \@ifxundefined [1]{%
 \@ifx{#1\undefined}
}%
\providecommand \@ifnum [1]{%
 \ifnum #1\expandafter \@firstoftwo
 \else \expandafter \@secondoftwo
 \fi
}%
\providecommand \@ifx [1]{%
 \ifx #1\expandafter \@firstoftwo
 \else \expandafter \@secondoftwo
 \fi
}%
\providecommand \natexlab [1]{#1}%
\providecommand \enquote  [1]{``#1''}%
\providecommand \bibnamefont  [1]{#1}%
\providecommand \bibfnamefont [1]{#1}%
\providecommand \citenamefont [1]{#1}%
\providecommand \href@noop [0]{\@secondoftwo}%
\providecommand \href [0]{\begingroup \@sanitize@url \@href}%
\providecommand \@href[1]{\@@startlink{#1}\@@href}%
\providecommand \@@href[1]{\endgroup#1\@@endlink}%
\providecommand \@sanitize@url [0]{\catcode `\\12\catcode `\$12\catcode
  `\&12\catcode `\#12\catcode `\^12\catcode `\_12\catcode `\%12\relax}%
\providecommand \@@startlink[1]{}%
\providecommand \@@endlink[0]{}%
\providecommand \url  [0]{\begingroup\@sanitize@url \@url }%
\providecommand \@url [1]{\endgroup\@href {#1}{\urlprefix }}%
\providecommand \urlprefix  [0]{URL }%
\providecommand \Eprint [0]{\href }%
\providecommand \doibase [0]{https://doi.org/}%
\providecommand \selectlanguage [0]{\@gobble}%
\providecommand \bibinfo  [0]{\@secondoftwo}%
\providecommand \bibfield  [0]{\@secondoftwo}%
\providecommand \translation [1]{[#1]}%
\providecommand \BibitemOpen [0]{}%
\providecommand \bibitemStop [0]{}%
\providecommand \bibitemNoStop [0]{.\EOS\space}%
\providecommand \EOS [0]{\spacefactor3000\relax}%
\providecommand \BibitemShut  [1]{\csname bibitem#1\endcsname}%
\let\auto@bib@innerbib\@empty
\bibitem [{\citenamefont {Qi}\ \emph {et~al.}(2008)\citenamefont {Qi},
  \citenamefont {Hughes},\ and\ \citenamefont {Zhang}}]{OMP1}%
  \BibitemOpen
  \bibfield  {author} {\bibinfo {author} {\bibfnamefont {X.-L.}\ \bibnamefont
  {Qi}}, \bibinfo {author} {\bibfnamefont {T.~L.}\ \bibnamefont {Hughes}},\
  and\ \bibinfo {author} {\bibfnamefont {S.-C.}\ \bibnamefont {Zhang}},\
  }\href@noop {} {\bibfield  {journal} {\bibinfo  {journal} {Physical Review
  B}\ }\textbf {\bibinfo {volume} {78}},\ \bibinfo {pages} {195424} (\bibinfo
  {year} {2008})}\BibitemShut {NoStop}%
\bibitem [{\citenamefont {Essin}\ \emph {et~al.}(2009)\citenamefont {Essin},
  \citenamefont {Moore},\ and\ \citenamefont {Vanderbilt}}]{OMP2}%
  \BibitemOpen
  \bibfield  {author} {\bibinfo {author} {\bibfnamefont {A.~M.}\ \bibnamefont
  {Essin}}, \bibinfo {author} {\bibfnamefont {J.~E.}\ \bibnamefont {Moore}},\
  and\ \bibinfo {author} {\bibfnamefont {D.}~\bibnamefont {Vanderbilt}},\
  }\href@noop {} {\bibfield  {journal} {\bibinfo  {journal} {Physical Review
  Letters}\ }\textbf {\bibinfo {volume} {102}},\ \bibinfo {pages} {146805}
  (\bibinfo {year} {2009})}\BibitemShut {NoStop}%
\bibitem [{\citenamefont {Malashevich}\ \emph {et~al.}(2010)\citenamefont
  {Malashevich}, \citenamefont {Souza}, \citenamefont {Coh},\ and\
  \citenamefont {Vanderbilt}}]{OMP3}%
  \BibitemOpen
  \bibfield  {author} {\bibinfo {author} {\bibfnamefont {A.}~\bibnamefont
  {Malashevich}}, \bibinfo {author} {\bibfnamefont {I.}~\bibnamefont {Souza}},
  \bibinfo {author} {\bibfnamefont {S.}~\bibnamefont {Coh}},\ and\ \bibinfo
  {author} {\bibfnamefont {D.}~\bibnamefont {Vanderbilt}},\ }\href@noop {}
  {\bibfield  {journal} {\bibinfo  {journal} {New Journal of Physics}\ }\textbf
  {\bibinfo {volume} {12}},\ \bibinfo {pages} {053032} (\bibinfo {year}
  {2010})}\BibitemShut {NoStop}%
\bibitem [{\citenamefont {Swiecicki}\ and\ \citenamefont {Sipe}(2014)}]{OMP4}%
  \BibitemOpen
  \bibfield  {author} {\bibinfo {author} {\bibfnamefont {S.~D.}\ \bibnamefont
  {Swiecicki}}\ and\ \bibinfo {author} {\bibfnamefont {J.~E.}\ \bibnamefont
  {Sipe}},\ }\href@noop {} {\bibfield  {journal} {\bibinfo  {journal} {Physical
  Review B}\ }\textbf {\bibinfo {volume} {90}},\ \bibinfo {pages} {125115}
  (\bibinfo {year} {2014})}\BibitemShut {NoStop}%
\bibitem [{\citenamefont {Essin}\ \emph {et~al.}(2010)\citenamefont {Essin},
  \citenamefont {Turner}, \citenamefont {Moore},\ and\ \citenamefont
  {Vanderbilt}}]{VanderbiltOMP}%
  \BibitemOpen
  \bibfield  {author} {\bibinfo {author} {\bibfnamefont {A.~M.}\ \bibnamefont
  {Essin}}, \bibinfo {author} {\bibfnamefont {A.~M.}\ \bibnamefont {Turner}},
  \bibinfo {author} {\bibfnamefont {J.~E.}\ \bibnamefont {Moore}},\ and\
  \bibinfo {author} {\bibfnamefont {D.}~\bibnamefont {Vanderbilt}},\
  }\href@noop {} {\bibfield  {journal} {\bibinfo  {journal} {Physical Review
  B}\ }\textbf {\bibinfo {volume} {81}},\ \bibinfo {pages} {205104} (\bibinfo
  {year} {2010})}\BibitemShut {NoStop}%
\bibitem [{\citenamefont {Vanderbilt}(2018)}]{BerryVanderbilt}%
  \BibitemOpen
  \bibfield  {author} {\bibinfo {author} {\bibfnamefont {D.}~\bibnamefont
  {Vanderbilt}},\ }\href@noop {} {\emph {\bibinfo {title} {Berry Phases in
  Electronic Structure Theory: Electric Polarization, Orbital Magnetization and
  Topological Insulators}}}\ (\bibinfo  {publisher} {Cambridge University
  Press},\ \bibinfo {year} {2018})\BibitemShut {NoStop}%
\bibitem [{\citenamefont {Resta}(1994)}]{Macro_Polar}%
  \BibitemOpen
  \bibfield  {author} {\bibinfo {author} {\bibfnamefont {R.}~\bibnamefont
  {Resta}},\ }\href@noop {} {\bibfield  {journal} {\bibinfo  {journal} {Rev.
  Mod. Phys}\ }\textbf {\bibinfo {volume} {66}},\ \bibinfo {pages} {899}
  (\bibinfo {year} {1994})}\BibitemShut {NoStop}%
\bibitem [{\citenamefont {Resta}(2010)}]{Elec_Polar}%
  \BibitemOpen
  \bibfield  {author} {\bibinfo {author} {\bibfnamefont {R.}~\bibnamefont
  {Resta}},\ }\href@noop {} {\bibfield  {journal} {\bibinfo  {journal} {J.
  Phys.: Condens. Matter}\ }\textbf {\bibinfo {volume} {22}},\ \bibinfo {pages}
  {123201} (\bibinfo {year} {2010})}\BibitemShut {NoStop}%
\bibitem [{\citenamefont {Xiao}\ \emph {et~al.}(2021)\citenamefont {Xiao},
  \citenamefont {Ren},\ and\ \citenamefont {Xiong}}]{SpinMagneto_Adiabatic}%
  \BibitemOpen
  \bibfield  {author} {\bibinfo {author} {\bibfnamefont {C.}~\bibnamefont
  {Xiao}}, \bibinfo {author} {\bibfnamefont {Y.}~\bibnamefont {Ren}},\ and\
  \bibinfo {author} {\bibfnamefont {B.}~\bibnamefont {Xiong}},\ }\href@noop {}
  {\bibfield  {journal} {\bibinfo  {journal} {Physical Review B}\ }\textbf
  {\bibinfo {volume} {103}},\ \bibinfo {pages} {115432} (\bibinfo {year}
  {2021})}\BibitemShut {NoStop}%
\bibitem [{\citenamefont {Mahon}\ \emph {et~al.}(2019)\citenamefont {Mahon},
  \citenamefont {Muniz},\ and\ \citenamefont {Sipe}}]{Perry_Sipe}%
  \BibitemOpen
  \bibfield  {author} {\bibinfo {author} {\bibfnamefont {P.~T.}\ \bibnamefont
  {Mahon}}, \bibinfo {author} {\bibfnamefont {R.~A.}\ \bibnamefont {Muniz}},\
  and\ \bibinfo {author} {\bibfnamefont {J.~E.}\ \bibnamefont {Sipe}},\
  }\href@noop {} {\bibfield  {journal} {\bibinfo  {journal} {Phys. Rev. B}\
  }\textbf {\bibinfo {volume} {99}},\ \bibinfo {pages} {235140} (\bibinfo
  {year} {2019})}\BibitemShut {NoStop}%
\bibitem [{\citenamefont {Malashevich}\ and\ \citenamefont
  {Souza}(2010)}]{OpticalConductivityIvo}%
  \BibitemOpen
  \bibfield  {author} {\bibinfo {author} {\bibfnamefont {A.}~\bibnamefont
  {Malashevich}}\ and\ \bibinfo {author} {\bibfnamefont {I.}~\bibnamefont
  {Souza}},\ }\href@noop {} {\bibfield  {journal} {\bibinfo  {journal}
  {Physical Review B}\ }\textbf {\bibinfo {volume} {82}},\ \bibinfo {pages}
  {245118} (\bibinfo {year} {2010})}\BibitemShut {NoStop}%
\bibitem [{\citenamefont {Mahon}\ and\ \citenamefont
  {Sipe}(2020{\natexlab{a}})}]{PerryOptical}%
  \BibitemOpen
  \bibfield  {author} {\bibinfo {author} {\bibfnamefont {P.~T.}\ \bibnamefont
  {Mahon}}\ and\ \bibinfo {author} {\bibfnamefont {J.~E.}\ \bibnamefont
  {Sipe}},\ }\href@noop {} {\bibfield  {journal} {\bibinfo  {journal} {Phys.
  Rev. Research 2}\ ,\ \bibinfo {pages} {043110}} (\bibinfo {year}
  {2020}{\natexlab{a}})}\BibitemShut {NoStop}%
\bibitem [{\citenamefont {Mahon}\ and\ \citenamefont
  {Sipe}(2020{\natexlab{b}})}]{PerryMagneto}%
  \BibitemOpen
  \bibfield  {author} {\bibinfo {author} {\bibfnamefont {P.~T.}\ \bibnamefont
  {Mahon}}\ and\ \bibinfo {author} {\bibfnamefont {J.~E.}\ \bibnamefont
  {Sipe}},\ }\href@noop {} {\bibfield  {journal} {\bibinfo  {journal} {Phys.
  Rev. Research 2}\ ,\ \bibinfo {pages} {033126}} (\bibinfo {year}
  {2020}{\natexlab{b}})}\BibitemShut {NoStop}%
\bibitem [{\citenamefont {Sipe}\ and\ \citenamefont
  {Ghahramani}(1993)}]{OpticalSipeGhahramani}%
  \BibitemOpen
  \bibfield  {author} {\bibinfo {author} {\bibfnamefont {J.~E.}\ \bibnamefont
  {Sipe}}\ and\ \bibinfo {author} {\bibfnamefont {E.}~\bibnamefont
  {Ghahramani}},\ }\href@noop {} {\bibfield  {journal} {\bibinfo  {journal}
  {Physical Review B}\ }\textbf {\bibinfo {volume} {48}},\ \bibinfo {pages}
  {11705} (\bibinfo {year} {1993})}\BibitemShut {NoStop}%
\bibitem [{\citenamefont {De~Groot}(1969)}]{MaxwellFields}%
  \BibitemOpen
  \bibfield  {author} {\bibinfo {author} {\bibfnamefont {S.~R.}\ \bibnamefont
  {De~Groot}},\ }\href@noop {} {\emph {\bibinfo {title} {The Maxwell Equations:
  Non Relativistic and Relativistic Derivation From Electron Theory (Studies in
  Statistical Mechanics)}}},\ Vol.~\bibinfo {volume} {IV}\ (\bibinfo
  {publisher} {North Holland},\ \bibinfo {year} {1969})\BibitemShut {NoStop}%
\bibitem [{Note1()}]{Note1}%
  \BibitemOpen
  \bibinfo {note} {For a review and references to original work see C.
  CohenTannoudji, J. Dupont-Roc, and G. Grynberg, Photons and Atoms.
  Introduction to Quantum Electrodynamics, John Wiley and Sons, Inc.,
  1989}\BibitemShut {NoStop}%
\bibitem [{\citenamefont {Healy}(1982{\natexlab{a}})}]{HealyPZW}%
  \BibitemOpen
  \bibfield  {author} {\bibinfo {author} {\bibfnamefont {W.}~\bibnamefont
  {Healy}},\ }\href@noop {} {\bibfield  {journal} {\bibinfo  {journal} {Phys.
  Rev. A}\ }\textbf {\bibinfo {volume} {26}},\ \bibinfo {pages} {1798}
  (\bibinfo {year} {1982}{\natexlab{a}})}\BibitemShut {NoStop}%
\bibitem [{\citenamefont {Foldy}\ and\ \citenamefont
  {Wouthuysen}(1950)}]{FoldyDirac}%
  \BibitemOpen
  \bibfield  {author} {\bibinfo {author} {\bibfnamefont {L.~L.}\ \bibnamefont
  {Foldy}}\ and\ \bibinfo {author} {\bibfnamefont {S.~A.}\ \bibnamefont
  {Wouthuysen}},\ }\href@noop {} {\bibfield  {journal} {\bibinfo  {journal}
  {Physical Review}\ }\textbf {\bibinfo {volume} {78}},\ \bibinfo {pages} {29}
  (\bibinfo {year} {1950})}\BibitemShut {NoStop}%
\bibitem [{\citenamefont {Bjorken}\ and\ \citenamefont
  {D.}(1964)}]{Bjorken_Drell}%
  \BibitemOpen
  \bibfield  {author} {\bibinfo {author} {\bibfnamefont {S.~D.}\ \bibnamefont
  {Bjorken}}\ and\ \bibinfo {author} {\bibfnamefont {D.~J.}\ \bibnamefont
  {D.}},\ }\href@noop {} {\emph {\bibinfo {title} {Relativistic Quantum
  Mechanics}}}\ (\bibinfo  {publisher} {McGraw-Hill},\ \bibinfo {year}
  {1964})\BibitemShut {NoStop}%
\bibitem [{\citenamefont {Schiff}(1968)}]{SchiffDirac}%
  \BibitemOpen
  \bibfield  {author} {\bibinfo {author} {\bibfnamefont {L.}~\bibnamefont
  {Schiff}},\ }\href@noop {} {\emph {\bibinfo {title} {Quantum Mechanics
  (International Pure and Applied Physics Series)}}}\ (\bibinfo  {publisher}
  {McGraw-Hill Companies},\ \bibinfo {year} {1968})\BibitemShut {NoStop}%
\bibitem [{\citenamefont {Xiao}\ \emph {et~al.}(2010)\citenamefont {Xiao},
  \citenamefont {Chang},\ and\ \citenamefont {Niu}}]{BerryPhase}%
  \BibitemOpen
  \bibfield  {author} {\bibinfo {author} {\bibfnamefont {D.}~\bibnamefont
  {Xiao}}, \bibinfo {author} {\bibfnamefont {M.-C.}\ \bibnamefont {Chang}},\
  and\ \bibinfo {author} {\bibfnamefont {Q.}~\bibnamefont {Niu}},\ }\href@noop
  {} {\bibfield  {journal} {\bibinfo  {journal} {Reviews of Modern Physics}\
  }\textbf {\bibinfo {volume} {82}},\ \bibinfo {pages} {1959} (\bibinfo {year}
  {2010})}\BibitemShut {NoStop}%
\bibitem [{\citenamefont {Iba$\tilde{\text{n}}$ez-Azpiroz}\ \emph
  {et~al.}(2015)\citenamefont {Iba$\tilde{\text{n}}$ez-Azpiroz}, \citenamefont
  {Eiguren}, \citenamefont {Bergara}, \citenamefont {Pettini},\ and\
  \citenamefont {Modugno}}]{Haldane_Julen}%
  \BibitemOpen
  \bibfield  {author} {\bibinfo {author} {\bibfnamefont {J.}~\bibnamefont
  {Iba$\tilde{\text{n}}$ez-Azpiroz}}, \bibinfo {author} {\bibfnamefont
  {A.}~\bibnamefont {Eiguren}}, \bibinfo {author} {\bibfnamefont
  {A.}~\bibnamefont {Bergara}}, \bibinfo {author} {\bibfnamefont
  {G.}~\bibnamefont {Pettini}},\ and\ \bibinfo {author} {\bibfnamefont
  {M.}~\bibnamefont {Modugno}},\ }\href@noop {} {\bibfield  {journal} {\bibinfo
   {journal} {Physical Review B}\ }\textbf {\bibinfo {volume} {92}},\ \bibinfo
  {pages} {195132} (\bibinfo {year} {2015})}\BibitemShut {NoStop}%
\bibitem [{\citenamefont {Kohn}\ and\ \citenamefont
  {Sham}(1965)}]{KohnSham-Ex}%
  \BibitemOpen
  \bibfield  {author} {\bibinfo {author} {\bibfnamefont {W.}~\bibnamefont
  {Kohn}}\ and\ \bibinfo {author} {\bibfnamefont {L.~J.}\ \bibnamefont
  {Sham}},\ }\href@noop {} {\bibfield  {journal} {\bibinfo  {journal} {Physical
  Review}\ }\textbf {\bibinfo {volume} {140}} (\bibinfo {year}
  {1965})}\BibitemShut {NoStop}%
\bibitem [{\citenamefont {Ogata}(2017)}]{OgataMagnetizationBloch}%
  \BibitemOpen
  \bibfield  {author} {\bibinfo {author} {\bibfnamefont {M.}~\bibnamefont
  {Ogata}},\ }\href@noop {} {\bibfield  {journal} {\bibinfo  {journal} {Journal
  of the Physical Society of Japan}\ }\textbf {\bibinfo {volume} {86}},\
  \bibinfo {pages} {044713} (\bibinfo {year} {2017})}\BibitemShut {NoStop}%
\bibitem [{\citenamefont {Ezawa}(2007)}]{QuantumFieldTheoretic}%
  \BibitemOpen
  \bibfield  {author} {\bibinfo {author} {\bibfnamefont {Z.~F.}\ \bibnamefont
  {Ezawa}},\ }\href@noop {} {\emph {\bibinfo {title} {Quantum Hall Effects:
  Field Theoretical Approach and Related Topics}}}\ (\bibinfo  {publisher}
  {World Scientific Publishing Company},\ \bibinfo {year} {2007})\BibitemShut
  {NoStop}%
\bibitem [{\citenamefont {Griffiths}(2013)}]{Griffiths}%
  \BibitemOpen
  \bibfield  {author} {\bibinfo {author} {\bibfnamefont {D.~J.}\ \bibnamefont
  {Griffiths}},\ }\href@noop {} {\emph {\bibinfo {title} {Introduction to
  Electrodynamics}}},\ \bibinfo {edition} {4th}\ ed.\ (\bibinfo  {publisher}
  {Pearson},\ \bibinfo {year} {2013})\BibitemShut {NoStop}%
\bibitem [{\citenamefont {Hodge}\ \emph {et~al.}(2014)\citenamefont {Hodge},
  \citenamefont {Migirditch},\ and\ \citenamefont {Kerr}}]{ElectronCurrent}%
  \BibitemOpen
  \bibfield  {author} {\bibinfo {author} {\bibfnamefont {W.~B.}\ \bibnamefont
  {Hodge}}, \bibinfo {author} {\bibfnamefont {S.~V.}\ \bibnamefont
  {Migirditch}},\ and\ \bibinfo {author} {\bibfnamefont {W.~C.}\ \bibnamefont
  {Kerr}},\ }\href@noop {} {\bibfield  {journal} {\bibinfo  {journal} {American
  Journal of Physics}\ }\textbf {\bibinfo {volume} {82}},\ \bibinfo {pages}
  {681} (\bibinfo {year} {2014})}\BibitemShut {NoStop}%
\bibitem [{\citenamefont {Cunha}\ \emph {et~al.}(2020)\citenamefont {Cunha},
  \citenamefont {Lima}, \citenamefont {Moraes}, \citenamefont {Fumeron},\ and\
  \citenamefont {Berche}}]{SpinCurrentNano}%
  \BibitemOpen
  \bibfield  {author} {\bibinfo {author} {\bibfnamefont {M.~M.}\ \bibnamefont
  {Cunha}}, \bibinfo {author} {\bibfnamefont {J.~R.~F.}\ \bibnamefont {Lima}},
  \bibinfo {author} {\bibfnamefont {F.}~\bibnamefont {Moraes}}, \bibinfo
  {author} {\bibfnamefont {S.}~\bibnamefont {Fumeron}},\ and\ \bibinfo {author}
  {\bibfnamefont {B.}~\bibnamefont {Berche}},\ }\href@noop {} {\bibfield
  {journal} {\bibinfo  {journal} {Journal of Physics: Condensed Matter}\
  }\textbf {\bibinfo {volume} {32}},\ \bibinfo {pages} {185301} (\bibinfo
  {year} {2020})}\BibitemShut {NoStop}%
\bibitem [{\citenamefont {Korani}\ and\ \citenamefont
  {Sabzyan}(2016)}]{SpinDynamics}%
  \BibitemOpen
  \bibfield  {author} {\bibinfo {author} {\bibfnamefont {Y.}~\bibnamefont
  {Korani}}\ and\ \bibinfo {author} {\bibfnamefont {H.}~\bibnamefont
  {Sabzyan}},\ }\href@noop {} {\bibfield  {journal} {\bibinfo  {journal}
  {Physical Chemistry Chemical Physics}\ ,\ \bibinfo {pages} {3166}} (\bibinfo
  {year} {2016})}\BibitemShut {NoStop}%
\bibitem [{\citenamefont {Chang}\ and\ \citenamefont
  {Niu}(2008)}]{BerryCurvOrb}%
  \BibitemOpen
  \bibfield  {author} {\bibinfo {author} {\bibfnamefont {M.-C.}\ \bibnamefont
  {Chang}}\ and\ \bibinfo {author} {\bibfnamefont {Q.}~\bibnamefont {Niu}},\
  }\href@noop {} {\bibfield  {journal} {\bibinfo  {journal} {Journal of
  Physics: Condensed Matter}\ }\textbf {\bibinfo {volume} {20}},\ \bibinfo
  {pages} {193202} (\bibinfo {year} {2008})}\BibitemShut {NoStop}%
\bibitem [{\citenamefont {Fisher}(1971)}]{MovingMagneticDipole}%
  \BibitemOpen
  \bibfield  {author} {\bibinfo {author} {\bibfnamefont {G.~P.}\ \bibnamefont
  {Fisher}},\ }\href@noop {} {\bibfield  {journal} {\bibinfo  {journal}
  {American Journal of Physics}\ }\textbf {\bibinfo {volume} {39}},\ \bibinfo
  {pages} {1528} (\bibinfo {year} {1971})}\BibitemShut {NoStop}%
\bibitem [{Note2()}]{Note2}%
  \BibitemOpen
  \bibinfo {note} {The Lagrangian of a single electron interacting with the
  electromagnetic field, which is specified in the laboratory frame, is then
  \begin {equation} \label {Lagrangian} \protect \text {L}=\protect \frac
  {1}{2}m\protect \textbf {v}\cdot \protect \textbf {v}+\protect \frac
  {e}{c}\protect \textbf {v}\cdot (\protect \textbf {A}+\protect \textbf
  {A}_\protect \text {static})+\protect \boldsymbol \mu \cdot \protect \textbf
  {E}_\protect \text {lattice}, \end {equation} together with contributions
  from the scalar potential and from the interaction of the magnetic dipole
  moment with the magnetic field. Since the last two terms on the
  right-hand-side of (\ref {Lagrangian}) both involve the velocity $\protect
  \textbf {v}$, they will both lead to a difference between the canonical
  momentum and the product of the mass and the velocity. Putting the spin
  magnetic moment $\protect \boldsymbol {\nu }=e\protect \textbf {S}/(mc)$,
  where $\protect \textbf {S}$ is the spin angular momentum, and including in
  this term the Thomas factor $\protect \frac {1}{2}$ to account for Thomas
  precession \cite {ThomasPrecession,JacksonEM}, we find \begin {equation}
  \protect \textbf {p}=m\protect \textbf {v}+\protect \frac {e}{c}(\protect
  \textbf {A}+\protect \textbf {A}_\protect \text {static})-\protect \frac
  {1}{2mc^2}(\protect \textbf {S}\times \nabla \protect \text {V}), \end
  {equation} Taking $\protect \textbf {S}\rightarrow {\protect \hbar \protect
  \boldsymbol \sigma }/2$, we can write this equation as \begin {equation}
  \label {spcden} e\protect \textbf {v}=\protect \frac {e}{m}(\protect \textbf
  {p}-\protect \frac {e}{c}(\protect \textbf {A}+\protect \textbf {A}_\protect
  \text {static}))+\protect \frac {e\protect \hbar }{4m^2c^2}(\protect
  \boldsymbol \sigma \times \nabla \protect \text {V}). \end {equation} The
  second term in the second line of the expression (\ref {cden}) for the
  current density is the field theory analogue of the last term on the
  right-hand-side of (\ref {spcden}).}\BibitemShut {Stop}%
\bibitem [{\citenamefont {Cohen}\ and\ \citenamefont
  {Louie}(2016)}]{CohenCondensedMatter}%
  \BibitemOpen
  \bibfield  {author} {\bibinfo {author} {\bibfnamefont {M.~L.}\ \bibnamefont
  {Cohen}}\ and\ \bibinfo {author} {\bibfnamefont {S.~G.}\ \bibnamefont
  {Louie}},\ }\href@noop {} {\emph {\bibinfo {title} {Fundamentals of Condensed
  Matter Physics}}}\ (\bibinfo  {publisher} {Cambridge University Press},\
  \bibinfo {year} {2016})\BibitemShut {NoStop}%
\bibitem [{\citenamefont {Blount}(1962)}]{SeitzSolid}%
  \BibitemOpen
  \bibfield  {author} {\bibinfo {author} {\bibfnamefont {E.}~\bibnamefont
  {Blount}},\ }\href@noop {} {\emph {\bibinfo {title} {Solid State Physics:
  Advances in Research and Applications}}},\ edited by\ \bibinfo {editor}
  {\bibfnamefont {F.}~\bibnamefont {Seitz}}\ and\ \bibinfo {editor}
  {\bibfnamefont {D.}~\bibnamefont {Turnbull}},\ Vol.~\bibinfo {volume} {13}\
  (\bibinfo  {publisher} {Bell Telephone Laboratories},\ \bibinfo {year}
  {1962})\ Chap.\ \bibinfo {chapter} {Formalisms of Band Theory}\BibitemShut
  {NoStop}%
\bibitem [{\citenamefont {Marzari}\ \emph {et~al.}(2012)\citenamefont
  {Marzari}, \citenamefont {Mostofi}, \citenamefont {Yates}, \citenamefont
  {Souza},\ and\ \citenamefont {Vanderbilt}}]{MaxLocWannier}%
  \BibitemOpen
  \bibfield  {author} {\bibinfo {author} {\bibfnamefont {N.}~\bibnamefont
  {Marzari}}, \bibinfo {author} {\bibfnamefont {A.~A.}\ \bibnamefont
  {Mostofi}}, \bibinfo {author} {\bibfnamefont {J.~R.}\ \bibnamefont {Yates}},
  \bibinfo {author} {\bibfnamefont {I.}~\bibnamefont {Souza}},\ and\ \bibinfo
  {author} {\bibfnamefont {D.}~\bibnamefont {Vanderbilt}},\ }\href@noop {}
  {\bibfield  {journal} {\bibinfo  {journal} {Reviews of Modern Physics}\
  }\textbf {\bibinfo {volume} {84}},\ \bibinfo {pages} {1419} (\bibinfo {year}
  {2012})}\BibitemShut {NoStop}%
\bibitem [{\citenamefont {Brouder}\ \emph {et~al.}(2007)\citenamefont
  {Brouder}, \citenamefont {Panati}, \citenamefont {Calandra}, \citenamefont
  {Mourougane},\ and\ \citenamefont {Marzari}}]{ExponentialLocWannier}%
  \BibitemOpen
  \bibfield  {author} {\bibinfo {author} {\bibfnamefont {C.}~\bibnamefont
  {Brouder}}, \bibinfo {author} {\bibfnamefont {G.}~\bibnamefont {Panati}},
  \bibinfo {author} {\bibfnamefont {M.}~\bibnamefont {Calandra}}, \bibinfo
  {author} {\bibfnamefont {C.}~\bibnamefont {Mourougane}},\ and\ \bibinfo
  {author} {\bibfnamefont {N.}~\bibnamefont {Marzari}},\ }\href@noop {}
  {\bibfield  {journal} {\bibinfo  {journal} {Physical Review Letters}\
  }\textbf {\bibinfo {volume} {98}},\ \bibinfo {pages} {046402} (\bibinfo
  {year} {2007})}\BibitemShut {NoStop}%
\bibitem [{\citenamefont {Matsuura}\ and\ \citenamefont
  {Ogata}(2016)}]{OgataCorrectionsPeierls}%
  \BibitemOpen
  \bibfield  {author} {\bibinfo {author} {\bibfnamefont {H.}~\bibnamefont
  {Matsuura}}\ and\ \bibinfo {author} {\bibfnamefont {M.}~\bibnamefont
  {Ogata}},\ }\href@noop {} {\bibfield  {journal} {\bibinfo  {journal} {Journal
  of the Physical Society of Japan}\ }\textbf {\bibinfo {volume} {85}},\
  \bibinfo {pages} {074709} (\bibinfo {year} {2016})}\BibitemShut {NoStop}%
\bibitem [{\citenamefont {Soluyanov}\ and\ \citenamefont
  {Vanderbilt}(2012)}]{SmoothGaugeVanderbilt}%
  \BibitemOpen
  \bibfield  {author} {\bibinfo {author} {\bibfnamefont {A.~A.}\ \bibnamefont
  {Soluyanov}}\ and\ \bibinfo {author} {\bibfnamefont {D.}~\bibnamefont
  {Vanderbilt}},\ }\href@noop {} {\bibfield  {journal} {\bibinfo  {journal}
  {Physical Review B}\ }\textbf {\bibinfo {volume} {85}},\ \bibinfo {pages}
  {115415} (\bibinfo {year} {2012})}\BibitemShut {NoStop}%
\bibitem [{\citenamefont {Winkler}\ \emph {et~al.}(2016)\citenamefont
  {Winkler}, \citenamefont {Soluyanov},\ and\ \citenamefont
  {Troyer}}]{SmoothGaugeWannier}%
  \BibitemOpen
  \bibfield  {author} {\bibinfo {author} {\bibfnamefont {G.~W.}\ \bibnamefont
  {Winkler}}, \bibinfo {author} {\bibfnamefont {A.~A.}\ \bibnamefont
  {Soluyanov}},\ and\ \bibinfo {author} {\bibfnamefont {M.}~\bibnamefont
  {Troyer}},\ }\href@noop {} {\bibfield  {journal} {\bibinfo  {journal}
  {Physical Review B}\ }\textbf {\bibinfo {volume} {93}},\ \bibinfo {pages}
  {035453} (\bibinfo {year} {2016})}\BibitemShut {NoStop}%
\bibitem [{\citenamefont {Panati}\ and\ \citenamefont
  {Pisante}(2013)}]{BlochBundlesMaxLoc}%
  \BibitemOpen
  \bibfield  {author} {\bibinfo {author} {\bibfnamefont {G.}~\bibnamefont
  {Panati}}\ and\ \bibinfo {author} {\bibfnamefont {A.}~\bibnamefont
  {Pisante}},\ }\href@noop {} {\bibfield  {journal} {\bibinfo  {journal}
  {Communications in Mathematical Physics}\ }\textbf {\bibinfo {volume}
  {322}},\ \bibinfo {pages} {835} (\bibinfo {year} {2013})}\BibitemShut
  {NoStop}%
\bibitem [{\citenamefont {Soluyanov}\ and\ \citenamefont
  {Vanderbilt}(2011)}]{Z2Insulator}%
  \BibitemOpen
  \bibfield  {author} {\bibinfo {author} {\bibfnamefont {A.~A.}\ \bibnamefont
  {Soluyanov}}\ and\ \bibinfo {author} {\bibfnamefont {D.}~\bibnamefont
  {Vanderbilt}},\ }\href@noop {} {\bibfield  {journal} {\bibinfo  {journal}
  {Physical Review B}\ }\textbf {\bibinfo {volume} {83}},\ \bibinfo {pages}
  {035108} (\bibinfo {year} {2011})}\BibitemShut {NoStop}%
\bibitem [{\citenamefont {Mayer}(2002)}]{Lowdin}%
  \BibitemOpen
  \bibfield  {author} {\bibinfo {author} {\bibfnamefont {I.}~\bibnamefont
  {Mayer}},\ }\href@noop {} {\bibfield  {journal} {\bibinfo  {journal}
  {International Jounal of Quantum Chemistry}\ }\textbf {\bibinfo {volume}
  {90}},\ \bibinfo {pages} {63} (\bibinfo {year} {2002})}\BibitemShut {NoStop}%
\bibitem [{\citenamefont {Healy}(1982{\natexlab{b}})}]{HealyQuantum}%
  \BibitemOpen
  \bibfield  {author} {\bibinfo {author} {\bibfnamefont {W.}~\bibnamefont
  {Healy}},\ }\href@noop {} {\emph {\bibinfo {title} {Non-Relativistic Quantum
  Electrodynamics}}}\ (\bibinfo  {publisher} {Academic Press},\ \bibinfo {year}
  {1982})\BibitemShut {NoStop}%
\bibitem [{\citenamefont {Woolley}(2020)}]{PZWNewPaper}%
  \BibitemOpen
  \bibfield  {author} {\bibinfo {author} {\bibfnamefont {R.~G.}\ \bibnamefont
  {Woolley}},\ }\href@noop {} {\bibfield  {journal} {\bibinfo  {journal}
  {Physical Review Research}\ }\textbf {\bibinfo {volume} {2}},\ \bibinfo
  {pages} {013206} (\bibinfo {year} {2020})}\BibitemShut {NoStop}%
\bibitem [{\citenamefont {Thonhauser}\ \emph {et~al.}(2005)\citenamefont
  {Thonhauser}, \citenamefont {Ceresoli}, \citenamefont {Vanderbilt},\ and\
  \citenamefont {Resta}}]{Magnetization_PI}%
  \BibitemOpen
  \bibfield  {author} {\bibinfo {author} {\bibfnamefont {T.}~\bibnamefont
  {Thonhauser}}, \bibinfo {author} {\bibfnamefont {D.}~\bibnamefont
  {Ceresoli}}, \bibinfo {author} {\bibfnamefont {D.}~\bibnamefont
  {Vanderbilt}},\ and\ \bibinfo {author} {\bibfnamefont {R.}~\bibnamefont
  {Resta}},\ }\href@noop {} {\bibfield  {journal} {\bibinfo  {journal}
  {Physical Review Letters}\ }\textbf {\bibinfo {volume} {95}},\ \bibinfo
  {pages} {137205} (\bibinfo {year} {2005})}\BibitemShut {NoStop}%
\bibitem [{\citenamefont {Ceresoli}\ \emph {et~al.}(2006)\citenamefont
  {Ceresoli}, \citenamefont {Thonhauser}, \citenamefont {Vanderbilt},\ and\
  \citenamefont {Resta}}]{Magnetization_MbI}%
  \BibitemOpen
  \bibfield  {author} {\bibinfo {author} {\bibfnamefont {D.}~\bibnamefont
  {Ceresoli}}, \bibinfo {author} {\bibfnamefont {T.}~\bibnamefont
  {Thonhauser}}, \bibinfo {author} {\bibfnamefont {D.}~\bibnamefont
  {Vanderbilt}},\ and\ \bibinfo {author} {\bibfnamefont {R.}~\bibnamefont
  {Resta}},\ }\href@noop {} {\bibfield  {journal} {\bibinfo  {journal} {Phys.
  Rev. B}\ }\textbf {\bibinfo {volume} {74}},\ \bibinfo {pages} {024408}
  (\bibinfo {year} {2006})}\BibitemShut {NoStop}%
\bibitem [{\citenamefont {Callaway}(1974)}]{CallawaySolidState}%
  \BibitemOpen
  \bibfield  {author} {\bibinfo {author} {\bibfnamefont {J.}~\bibnamefont
  {Callaway}},\ }\href@noop {} {\emph {\bibinfo {title} {Quantum Theory of the
  Solid State: Part A}}}\ (\bibinfo  {publisher} {Academic Press New York and
  London},\ \bibinfo {year} {1974})\BibitemShut {NoStop}%
\bibitem [{\citenamefont {Manoukian}(2006)}]{ManoukianBook}%
  \BibitemOpen
  \bibfield  {author} {\bibinfo {author} {\bibfnamefont {E.}~\bibnamefont
  {Manoukian}},\ }\href@noop {} {\emph {\bibinfo {title} {Quantum Theory: A
  Wide Spectrum}}}\ (\bibinfo  {publisher} {Springer},\ \bibinfo {year}
  {2006})\BibitemShut {NoStop}%
\bibitem [{\citenamefont {Dresselhaus}\ \emph {et~al.}(2008)\citenamefont
  {Dresselhaus}, \citenamefont {Dresselhaus},\ and\ \citenamefont
  {Jorio}}]{GroupTheoryDresselhaus}%
  \BibitemOpen
  \bibfield  {author} {\bibinfo {author} {\bibfnamefont {M.~S.}\ \bibnamefont
  {Dresselhaus}}, \bibinfo {author} {\bibfnamefont {G.}~\bibnamefont
  {Dresselhaus}},\ and\ \bibinfo {author} {\bibfnamefont {A.}~\bibnamefont
  {Jorio}},\ }\href@noop {} {\emph {\bibinfo {title} {Group Theory: Application
  to the physics of Condensed Matter}}}\ (\bibinfo  {publisher} {Springer},\
  \bibinfo {year} {2008})\BibitemShut {NoStop}%
\bibitem [{\citenamefont {Thomas}(1927)}]{ThomasPrecession}%
  \BibitemOpen
  \bibfield  {author} {\bibinfo {author} {\bibfnamefont {L.}~\bibnamefont
  {Thomas}},\ }\href@noop {} {\bibfield  {journal} {\bibinfo  {journal} {The
  London, Edinburgh, and Dublin Philosophical Magazine and Journal of Science}\
  }\textbf {\bibinfo {volume} {3}},\ \bibinfo {pages} {1} (\bibinfo {year}
  {1927})}\BibitemShut {NoStop}%
\bibitem [{\citenamefont {Jackson}(1999)}]{JacksonEM}%
  \BibitemOpen
  \bibfield  {author} {\bibinfo {author} {\bibfnamefont {J.~D.}\ \bibnamefont
  {Jackson}},\ }\href@noop {} {\emph {\bibinfo {title} {Classical
  Electrodynamics}}}\ (\bibinfo  {publisher} {John Wiley and Sons},\ \bibinfo
  {year} {1999})\BibitemShut {NoStop}%
\end{thebibliography}%
\end{document}